\documentclass[twocolumn,twocolappendix,trackchanges]{aastex631}
\hypersetup{linkcolor=red,citecolor=green,filecolor=cyan,urlcolor=magenta}

\shorttitle{Hot core chemistry in the outer Galaxy} 
\shortauthors{T. Ikeda et al.} 

\usepackage{amsmath}
\usepackage{amsfonts}
\usepackage{graphicx}
\usepackage{rotating}
\usepackage{txfonts}
\usepackage{color}
\usepackage{wrapfig}
\usepackage{amssymb}
\usepackage{color}

\begin{document}

%\title{} 
%\title{Digging into the chemical complexity in low-metallicity environments: \\An observational study of a hot molecular core in the outer Galaxy} 
\title{Digging into the chemical complexity in the outer Galaxy: A hot molecular core in Sharpless 2-283} 

%%%%%%%%%%%%%%%%%%%%%%%%%%%%%%%%%%%%%%%%%%%%
%1. Digging into the chemical complexity in low-metallicity environments: New detection of a hot molecular core in the outer Galaxy

%2. How environmental condition affects the molecular evolution in star-forming regions: An observational study of hot core chemistry in the outer Galaxy

%3. The impact of low-metallicity on astrochemistry: An observational study of hot molecular core in the outer Galaxy

%4. Astrochemical Insights into low-metallicity environments: A new detection of  hot molecular core in the outer Galaxy

%5. The detection of hot molecular core in the outer Galaxy
%%%%%%%%%%%%%%%%%%%%%%%%%%%%%%%%%%%%%%%%%%%%

\correspondingauthor{Toki Ikeda} 
\email{tokiikeda0122@gmail.com}

\correspondingauthor{Takashi Shimonishi} 
\email{shimonishi@env.sc.niigata-u.ac.jp} 

\author[0009-0004-5964-3892]{Toki Ikeda} 
\affiliation{Department of Natural Environmental Science, Graduate School of Science and Technology, Niigata University, Ikarashi-ninocho 8050, Nishi-ku, Niigata 950-2181, Japan}

\author[0000-0002-0095-3624]{Takashi Shimonishi} 
\affiliation{Institute of Science and Technology, Niigata University, Ikarashi-ninocho 8050, Nishi-ku, Niigata 950-2181, Japan}

\author[0000-0002-2699-4862]{Hiroyuki Kaneko}
\affiliation{College of Creative Studies, Niigata University,
8050 Ikarashi 2-no-cho, Nishi-ku, Niigata, Niigata 950-2181, Japan}

\author[0000-0002-2026-8157]{Kenji Furuya} 
\affiliation{Department of Astronomy, Graduate School of Science, University of Tokyo, Tokyo 113-0033, Japan}
\affiliation{RIKEN Pioneering Research Institute, 2-1 Hirosawa, Wako-shi, Saitama 351-0198, Japan}

\author[0000-0002-6907-0926]{Kei E. I. Tanaka} 
\affiliation{Department of Earth and Planetary Sciences, Institute of Science Tokyo, Meguro, Tokyo, 152-8551, Japan}

\author[0000-0003-1604-9127]{Natsuko Izumi} 
\affiliation{National Astronomical Observatory of Japan, 2-21-1 Osawa, Mitaka, Tokyo 181-8588, Japan}

\begin{abstract}
%Comprehending interstellar chemistry in low-metallicity environments is crucial for understanding early cosmic chemical evolution. 
The outer Galaxy (galactocentric distance $\gtrsim$13.5 kpc) serves as an excellent laboratory for investigating the chemical complexity in low-metallicity environments. 
Here, we present the chemical analyses for the outer Galactic hot core Sh 2-283-1a SMM1 ($D_\mathrm{GC}$ = 15.7 kpc and $Z$ $\sim$0.3 $Z_\odot$), recently detected by \citet{Ikeda2025} using ALMA.
Toward this source, a variety of molecular species, including complex organic molecules (COMs: CH$_3$OH, $^{13}$CH$_3$OH, CH$_2$DOH, and CH$_3$OCH$_3$) are detected.
%The molecular abundances relative to CH$_3$OH are similar to those of another outer Galactic hot core, suggesting that such chemically rich hot cores are ubiquitous in the outer Galaxy.
The molecular abundances relative to CH$_3$OH are similar to those of another outer Galactic hot core, demonstrating that chemically rich hot cores exist in different regions of the outer Galaxy.
We also compared molecular abundances among hot cores in the inner Galaxy, outer Galaxy, and Magellanic Clouds.
This comparison revealed that the metallicity-corrected $N$(SO$_2$)/$N$(H$_2$) ratios of outer Galactic hot cores are significantly lower than those of the inner Galactic ones, while their $N$(CH$_3$OH)/$N$(H$_2$) ratios are similar.
The Magellanic hot cores show different trends despite having metallicities similar to those of the outer Galaxy, indicating that the chemical complexity of hot cores is governed by environmental conditions (e.g., cosmic ray intensity and dust temperature) rather than simple metallicity scaling.
These environmental differences would also affect the production efficiency of COMs derived from CH$_3$OH, as the $N$(CH$_3$OCH$_3$)/$N$(CH$_3$OH) and $N$(C$_2$H$_5$OH)/$N$(CH$_3$OH) ratios in the outer Galactic sources are moderately lower than those of inner Galactic sources.
The $N$(CH$_2$DOH)/$N$(CH$_3$OH) ratio of Sh 2-283-1a SMM1 is 1.5$^{+3.9}_{-1.2}$$\%$, comparable to that of inner Galactic high-mass sources.

%Comparison with another outer Galactic hot core revealed remarkable similarity in molecular abundances relative to CH3OH, suggesting universal chemical complexity in the outer Galaxy. While metallicity-scaled N(CH3OH)/N(H2) ratios of outer Galactic hot cores resemble inner Galactic ones, their N(SO2)/N(H2) ratios are over an order of magnitude lower. However, N(CH3OCH3; C2H5OH)/N(CH3OH) ratios are similar across regions, indicating consistent COM formation efficiency once CH3OH is present. Furthermore, the CH2DOH/CH3OH ratio (1.4 ± 0.9 %) suggests efficient deuterium fractionation occurred under cold conditions (≲ 20 K) during CH3OH formation, comparable to inner Galactic intermediate to high-mass sources. These findings collectively indicate that hot core chemical complexity cannot simply be scaled with metallicity.
\end{abstract}

\keywords{astrochemistry --- ISM: molecules --- stars: protostars --- outer galaxy --- radio lines: ISM}

\section{Introduction} \label{sec_intro} 
%The outer Galaxy, which is defined as having a galactocentric distance ($D_{\rm{GC}}$) larger than $\sim$13.5 kpc, is an excellent laboratory for understanding the chemical complexity in different environments from the solar-neighborhood.
The outer Galaxy, which is defined as having a galactocentric distance ($D_{\rm{GC}}$) larger than $\sim$13.5 kpc, where the gas surface density starts to drop \citep[e.g.,][]{Wolfire_2003, Sun2024}, is an excellent laboratory for understanding the chemical complexity in different environments from the solar-neighborhood.
This region presents the distinctive physical properties, such as lower metallicity compared to the solar-neighborhood \citep[$\lesssim$1/3 $Z_\odot$;][]{Fer17, Wenger2019} and lower gas density \citep{Nak16}.
Moreover, perturbations from spiral arms and supernova explosions are expected to be minimal, as this region lies beyond the main arms of the Galaxy and exhibits a significantly low star formation rate \citep[e.g., ][]{Reid2019, Elia2022}.
In addition, its proximity to the Sun allows us to investigate more compact structures in star-forming regions than in the other known low-metallicity environments, such as the Large Magellanic Cloud (LMC; $\sim1/2$--$1/3\:Z_\odot$) and the Small Magellanic Cloud (SMC; $\sim1/5$--$1/10\:Z_\odot$), which are located at distances of $\sim50$--$60$ kpc \citep{Pie13, Graczyk2020}.

A hot molecular core (hot core) represents one of the early phases in star formation, playing a pivotal role in the emergence and development of complex molecules in the interstellar medium (ISM).
Hot cores are characterized by the presence of warm ($T \gtrsim$100 K), dense ($n$ $\gtrsim$10$^6$ cm$^{-3}$), and compact ($<$0.1 pc) molecular gas associated with a central protostar \citep[e.g.,][]{van1998, Kurtz2000}.
The chemically rich nature of hot cores is initiated by the sublimation of ice mantles formed on the dust surfaces.
Before and during the heating of ice mantles by stellar radiation, grain surface reactions are thought to contribute to the formation of complex organic molecules (COMs: molecules consisting of at least six atoms) \citep[e.g.,][]{Gar06, Herbst2009}. 
These processes enrich gas-phase molecules, and consequently, various molecular species can be detected towards hot cores with radio and infrared observations via molecular transitions. 
Hence, the detailed study of hot core chemistry is crucial to unveil the chemical evolution in star-forming regions.

%This process causes the enrichment of the gas-phase molecules, including complex organic molecules (COMs: molecules consisting of at least six atoms) and their precursor species such as CH$_3$OH and NH$_3$. 
%In hot gas ($T >$100 K), COMs larger than CH$_3$OH are efficiently formed via the gas-phase reactions \citep[e.g.,][]{Nomura2004, Gar06, Her09}.
%As a result, a variety of molecular species, in particular COMs, can be detected towards hot cores with radio observations as rotational transition lines of molecules.

%ホットコアchemistryの環境依存性の話
%LMC, SMCの話
%銀河系全体における普遍性 (特に外縁部の話)
%外縁部のサンプルを増やすために原始星サーベイを実施した話 (成果も含めて)
%new detection of hot coreの話、主題を含めて
%論文の構成

Recent ALMA (the Atacama Large Millimeter/submillimeter Array) observations have revealed the presence of hot cores in low-metallicity environments, such as the LMC and SMC \citep{Shimonishi2016, Shimonishi2020, Shimonishi2023,Shimonishi_magos, Sewilo2018, Sewilo2022, Golshan2024, Broadmeadow2025}, and the outer Galaxy at $D_{\rm{GC}}$ of 19.0 kpc \citep[$\sim$1/4 $Z_\odot$;][]{Shimonishi2021}.
%These regions are known to have low-metallicity environments compared to the solar neighborhood.
The comparison of molecular abundances among the Magellanic hot cores has suggested that there are large variations, especially for organic molecules.
In some sources, organic molecules are significantly depleted beyond what would be expected from metallicity differences alone \citep[see ][and references therein]{ST24}. 
This large variation suggests that the molecular compositions of hot cores do not simply scale with the environmental metallicity, but are also strongly influenced by other factors, such as the degree of surrounding star-formation activity.
On the other hand, detections of hot cores in the outer Galaxy have so far been limited, with only one source reported in \cite{Shimonishi2021}. 
They have reported that the outer Galactic hot core, WB89-789 SMM1, shows a very similar organic molecular composition to its counterparts in the inner Galaxy. 
However, it remains unclear whether such molecular complexity is also present in other star-forming regions in the outer Galaxy. 
Although the outer Galaxy and the Magellanic Clouds share similar low-metallicity conditions, their environments are expected to differ significantly due to the differences in star formation activity. 
The Magellanic Clouds are characterized by active star formation, which presumably results in strong stellar feedback, such as high cosmic ray intensity and high dust temperature \citep[e.g.,][]{Shimonishi2016,Shimonishi2020}. 
In contrast, the outer Galaxy is a relatively quiescent environment where such external influence would be much weaker. 
Since photochemistry and dust surface chemistry play a vital role in generating complex molecules in star-forming regions \citep[e.g., ][and references therein]{Herbst2009}, comparing the chemical properties of hot cores in these regions allows us to investigate whether the chemical properties vary depending on environmental factors other than metallicity (e.g., cosmic ray intensity).
Therefore, increasing the sample size of hot cores in the outer Galaxy is crucial for such comparative studies.
%To understand molecular evolution in relatively quiescent star-forming environments, which differ from the Magellanic Clouds but share similarly low metallicity conditions, surveys of hot cores in the outer Galaxy have been necessary. 
%This would indicate that the chemical properties of hot cores in the LMC do not solely depend on the metallicity but on other environmental factors such as, e.g.,  dust temperature during the initial ice-forming stage \citep{Acharyya2018, Shimonishi2020}.
%On the other hand, the outer Galactic hot core, WB89-789 SMM1, showed a remarkable similarity in its chemical abundances relative to CH$_3$OH when compared with the inner Galactic hot core \citep{Shimonishi2021}.
%This result indicates that the metallicity-scaled chemistry exists in the outer Galactic hot core, although at the time of that study, this was the only hot core detected in the outer Galaxy.

%To increase the number of hot core samples and understand the chemical complexity of hot cores in the outer Galaxy, 
We conducted a protostar survey toward the five outer Galactic star-forming regions (Sh 2-283, NOMF05-16, 19, 23, and 63; $D_{\rm{GC}}$ = 15.7-17.4 kpc) with ALMA.
Detailed information about the survey, including target selection criteria, can be found in \cite{Ikeda2025}. 
As a result of the survey, five protostellar objects associated with bipolar outflows were detected with CO($J$ = 3--2) line emission, indicating that star formation is ongoing in these regions \citep{Ikeda2025}.
Moreover, one of these outflow-driving sources, Sh 2-283-1a SMM1, showed a chemically rich nature, and a variety of molecular transition lines including high-excitation lines ($E_u >$100 K) of CH$_3$OH and SO$_2$ were detected.

In this paper, we report the results of the detailed chemical analyses for this hot core located in the outer Galaxy.
In Section \ref{sec_tarobsred}, we describe the observation details and the data reduction.
The observational results, including the observed molecular spectra and images, as well as analyses of the physical and chemical properties of the source, are given in Section \ref{sec_res}.
Discussions on the comparisons of the chemical properties of this source with the inner Galactic and Magellanic hot cores are presented in Section \ref{sec_disc}.
The summary of this paper is given in Section \ref{sec_sum}.

\section{Observations and data reduction} \label{sec_tarobsred} 
\subsection{Sh 2-283-1a SMM1} \label{sec_tar} 
The target source is located in the star-forming region Sh 2-283 in the outer Galaxy and is known to be associated with an H$_\mathrm{II}$ region \citep{Fich1991}.
%This region is a known H$_\mathrm{II}$ region located in the outer Galaxy \citep{Fich1991}.
%Based on the model A5 in \cite{Reid2014}, the kinematic and galactocentric distances of Sh 2-283 are estimated to be 7.2 kpc and 15.2 kpc, respectively.
%They were derived from model A5 in \citep{Reid2014}.
According to the astrometric data of a nearby cluster member (ALS 18674) in the Gaia Early Data Release 3 \citep[Gaia source ID 3119827723711464576,][]{Gaia2021,Bra19}, the distance to the Sh 2-283 region is estimated to be 7.9$^{+1.2}_{-1.1}$ kpc, which corresponds to the galactocentric distance of 15.7$^{+1.1}_{-0.9}$ kpc \citep{Ikeda2025}.  
%The distance was directly derived from the parallax (0.1269759 mas), and its parallax bias was corrected following the method described in \cite{Lindegren2021}.
The metallicity of this region is estimated to be approximately 30$\%$ of the solar value based on optical spectroscopy of the associated H$_\mathrm{II}$ region \citep{Fer17}.

%Additional four star-forming regions \citep[NOMF05-16, 19, 23, and 63; ][]{Nakagawa2005} were separately covered in our ALMA observations.
%Detailed information about these sources, including target selection criteria, can be found in \cite{Ikeda2025}.
%In this paper, we focus on Sh 2-283-1a SMM1, which shows a chemically rich nature. 

\subsection{Observations} \label{sec_obs} 
All data used in this work were obtained by ALMA as a part of the Cycle 9 program (2022.1.01270.S, PI: T. Shimonishi).
%The detailed description about the observations for all target sources can be found in \cite{Ikeda2025}.
The band 7 receiver was used to observe molecular emission lines (such as $^{12}$CO($J$ = 3--2), HCO$^+$($J$ = 4--3), SiO($J$ = 8--7), SO, SO$_2$, and CH$_3$OH) and 0.87 mm continuum.
Five spectral windows (SPWs) of the correlator were used in the frequency division mode.
The channel spacing for all SPWs was set to 976.6 kHz (corresponding to 0.85 km s$^{-1}$ at 345 GHz).
Two SPWs with a bandwidth of 0.938 GHz cover the sky frequencies of 344.038--344.976 and 344.978--345.916 GHz.
The other three SPWs were set to cover the sky frequencies of 345.919--347.794, 346.169--358.044, and 356.961--358.836 GHz, in which the bandwidth was 1.875 GHz.
Sh 2-283-1a SMM1 was observed in October 2022.
The antenna configuration was C-2, resulting in a maximum baseline length of 313 m.
The on-source time was 27.2 minutes.
The observations were conducted in three execution blocks.
During the observation period, 45 antennas were operated except for one observation block, in which 42 antennas were used.
The bandpass calibrators were J0510+1800 and J0750+1231, and the phase calibrator was J0641-0320.

\subsection{Data reduction} \label{sec_red} 
Data reduction, including calibration and imaging, was performed with the Common Astronomy Software Applications package  \citep[CASA: ][]{CASA2022} version 6.4.1.12.
The raw data were calibrated with the pipeline script provided by the observatory for initial data flagging, the calibration of bandpass, complex gain, and flux scaling (pipeline version 2022.2.0.64).
The visibilities of the line emissions were obtained by subtracting the continuum visibilities using the {\itshape uvcontsub} task in CASA.
The clean images were produced using the {\itshape tclean} task in CASA. 
The line image cubes were obtained adopting Briggs weighting with a robust parameter of 0.5 for each SPW.
%Since a variety of molecular emissions were detected toward Sh 2-283-1a SMM1, 
In order to improve the signal-to-noise ratio, we performed phase self-calibration using the continuum after the calibration.
One round of phase self-calibration was applied with a solution interval of 60 seconds.
Finally, we corrected the primary beam patterns using the {\itshape impbcor} task.
The systematic error on the absolute flux was approximately 10\% according to the ALMA technical handbook \footnote{\url{https://almascience.nao.ac.jp/documents-and-tools/cycle11/alma-technical-handbook}}.

The average beam size of the line images is 0$\farcs$79 $\times$ 0$\farcs$68, which corresponds to 6.2 $\times$ 10$^3$ au $\times$ 5.4 $\times$ 10$^3 $ au at the distance of Sh 2-283 (7.9 kpc from the Sun).
The maximum recoverable scale is about 6$\farcs$3 ($\sim$5.0 $\times$ 10$^4$ au at 7.9 kpc).
%\edit1{The channel spacing for all SPWs was set to 976.6 kHz (corresponding to 0.85 km s$^{-1}$ at 345 GHz).}

\section{Results and analysis} \label{sec_res} 
\subsection{Spectra} \label{sec_spc}  
\begin{figure*}[tp!]
\begin{center}
\includegraphics[width=13.0cm]{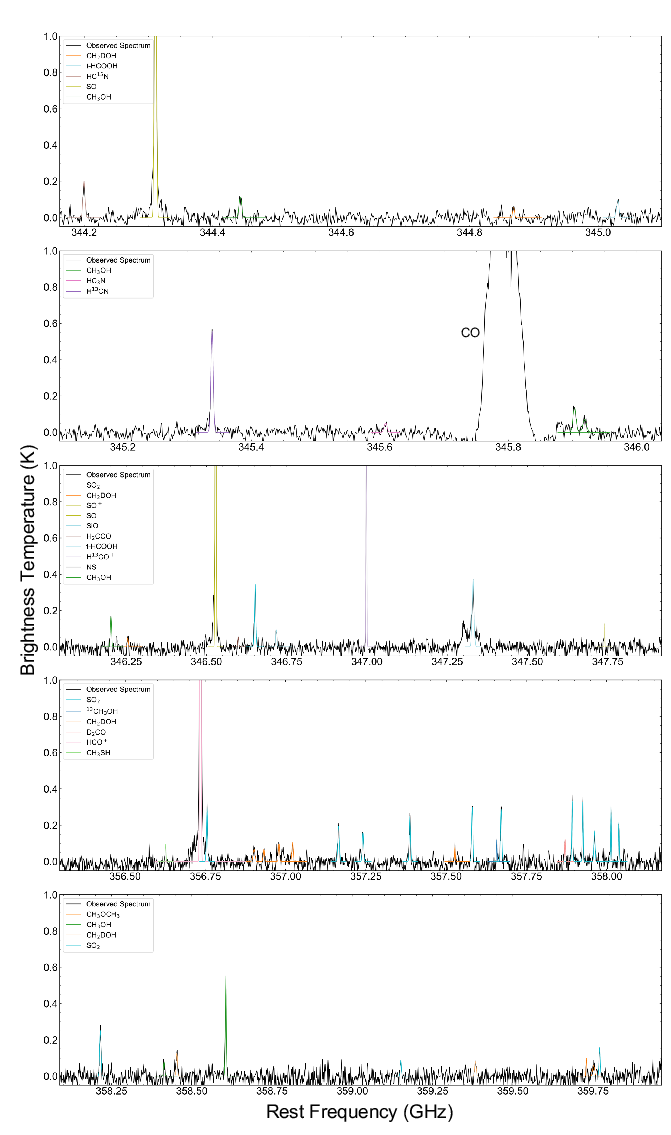}
\caption{ALMA Band 7 spectra of Sh 2-283-1a SMM1 extracted from the elliptical region (0$\farcs$89 $\times$ 0$\farcs$70) centered at RA = 6$^\mathrm{h}$38$^\mathrm{m}$29$\fs$660 and Dec = 0$\arcdeg$44$\arcmin$40$\farcs$75 (ICRS), which corresponds to the emission peak of the 0.87 mm continuum and molecular emissions.
The lines with different colors represent the results of Gaussian fitting for each molecular line (Table \ref{tab_line}).
%The detected molecular lines are labeled.
The systemic velocity of 53.2 km s$^{-1}$ is assumed to identify the molecular lines.}
\label{im_spec}
\end{center}
\end{figure*}

Figure \ref{im_spec} shows the submillimeter spectra extracted from the elliptical region (0$\farcs$89 $\times$ 0$\farcs$70) centered at RA = 6$^\mathrm{h}$38$^\mathrm{m}$29$\fs$660 and Dec = 0$\arcdeg$44$\arcmin$40$\farcs$75 (ICRS), which corresponds to the emission peak of CH$_3$OH and the 0.87 mm continuum.
%With determining the systematic velocity ($V_\mathrm{{sys}}$) as 53.2 km s$^{-1}$ from CH$_3$OH (5$_{4}$ A$^-$--6$_{3}$ A$^-$) line, spectral lines are identified with the aid of Cologne Database for Molecular Spectroscopy 
Spectral lines were identified by determining the systemic velocity ($V_\mathrm{{sys}}$) as 53.2 km s$^{-1}$ from the CH$_3$OH (5$_{4}$ A$^-$--6$_{3}$ A$^-$) line, with the aid of the Cologne Database for Molecular Spectroscopy \citep[CDMS,][]{Muller2001, Muller2005} and the Jet Propulsion Laboratory \citep[JPL,][]{Pickett1998}. 
The detection criteria adopted here were a 3$\sigma$ significance level and the velocity coincidence of the line emission with $V_\mathrm{{sys}}$ of the target ($\sim$$\pm$2 km s$^{-1}$ from $V_\mathrm{{sys}}$).
For the lines whose significance level was higher than 2.5$\sigma$ but lower than 3$\sigma$ and the emission distribution coincides with the continuum peak, we regarded them as tentative detections.
We derived the line parameters (the peak brightness temperature, FWHM, LSR velocity, and integrated intensity) by fitting a Gaussian profile to each detected line (colored lines in Figure \ref{im_spec}).
%For some weak lines, two channels were binned, or the Savitzky–Golay filter was used to reduce noise.
For some weak lines, two channels were binned, or the Savitzky-Golay filter (window size of 5 channels and polynomial order of 3) was used to reduce noise.
The line parameters for the detected molecular lines are summarized in Table \ref{tab_line}.
The detailed fitting results are summarized in Appendix \ref{sec_appen_fit}.

As shown in Figure \ref{im_spec}, we detected a variety of molecular species such as CO, SO, SO$^+$, SO$_2$, HCO$^+$, H$^{13}$CO$^+$, SiO, H$^{13}$CN, HC$^{15}$N, D$_2$CO, \textit{trans}-HCOOH, CH$_3$OH, $^{13}$CH$_3$OH, CH$_2$DOH, and CH$_3$OCH$_3$ toward Sh 2-283-1a SMM1.
For SO$_2$, CH$_3$OH, and CH$_2$DOH, multiple high excitation lines with different upper state energies were detected.
We also tentatively detected NS, H$_2$CCO, HC$_3$N, and CH$_3$SH.
Although CH$_3$SH (14$_{1,13,1}$--13$_{1,12,1}$) was detected with a signal above 3$\sigma$ of the rms noise level (see Table \ref{tab_line}), the other CH$_3$SH line (14$_{1,13,0}$--13$_{1,12,0}$), which has a similar upper state energy and line intensity ($E_{\rm{u}}$ = 133.8 K), remained undetected.
We thus regard the CH$_3$SH detection as tentative.

The typical measured line widths are 3--6 km s$^{-1}$.
Although most of the molecular lines show a single velocity component peaked near $V_{\rm{sys}}$, CO has multiple velocity components because it traces protostellar outflows and jets \citep{Ikeda2025}.
Note that the derived intensities of the lines associated with broad wing components (e.g., SO and HCO$^+$) may be slightly overestimated due to the contamination from the broad components.

\begin{figure*}[tp!]
\begin{center}
\includegraphics[width=16.0cm]{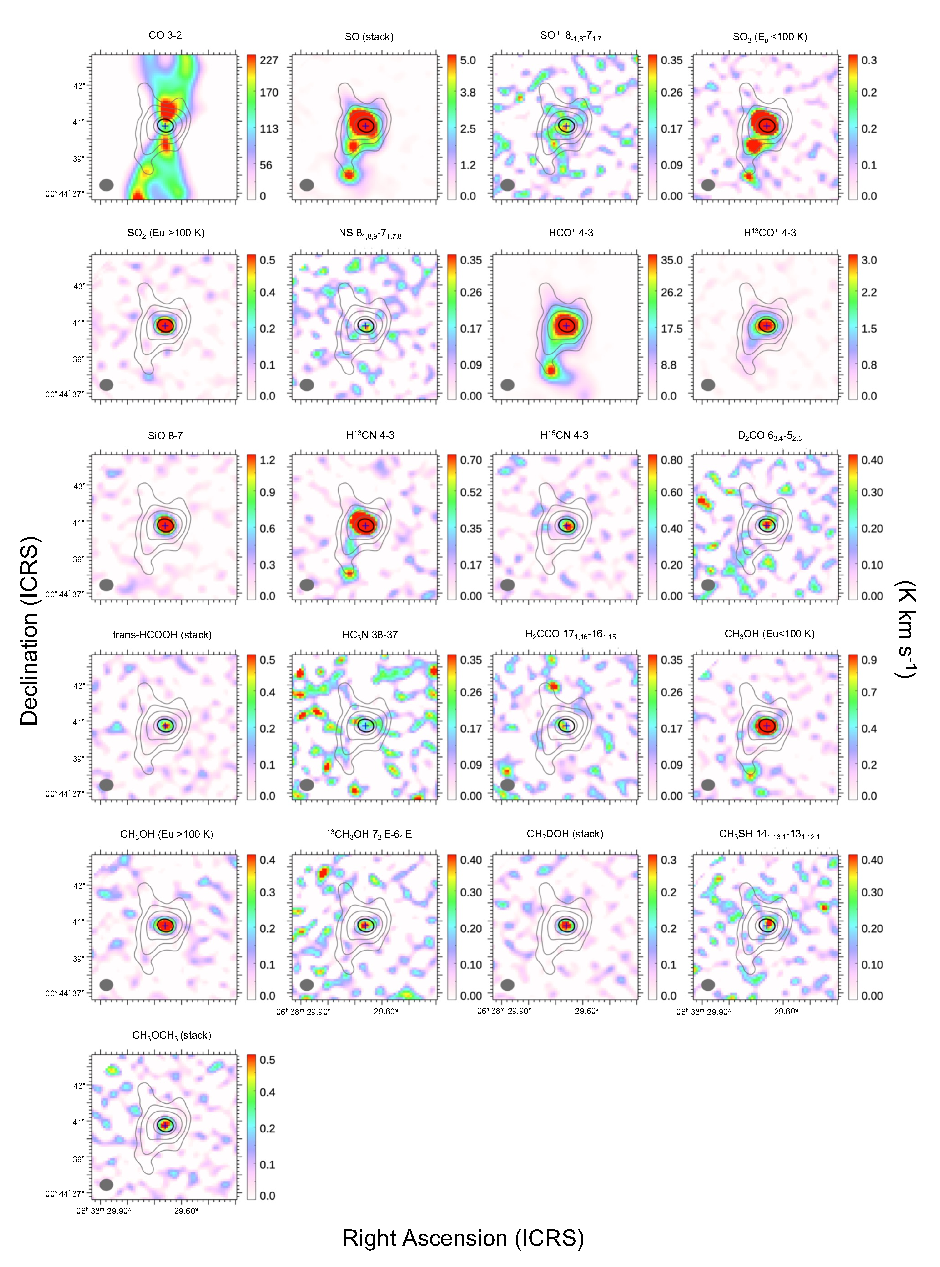}
\caption{The integrated intensity maps of the detected molecular lines.
Grey contours show the emission distribution of 0.87 mm  continuum, and the contour levels are 5$\sigma$, 15$\sigma$, 25$\sigma$, and 100$\sigma$ of the rms noise level (1$\sigma$ = 0.07 mJy beam$^{-1}$).
The blue cross indicates the peak position of 0.87 mm continuum.
The black open circle represents the extraction region of the spectra used for the discussion in the text.
%North is up, and east is to the left.
For molecular species with multiple line detections, lines are stacked in order to reduce the noise level.
The synthesized beam size is shown in the bottom left corner, as a gray ellipse.
%The black allow in SO panel points the position of the secondly peak.
}
\label{im_map}
\end{center}
\end{figure*}

\subsection{Images} \label{sec_img} 
Figure \ref{im_map} shows the integrated intensity maps of the detected molecular species.
The images were generated by integrating the spectral data in the velocity range where the emission is detected.
The integration range (typically $\sim$$\pm$FWHM) was determined by visual inspection for each line to ensure that broad wing components were included.

For SO, SO$_2$, \textit{trans}-HCOOH, CH$_3$OH, CH$_2$DOH, and CH$_3$OCH$_3$, several lines were stacked to reduce the noise level.
Contours and color in this figure show the 0.87 mm continuum and each molecular emission, respectively.

Most molecular emissions have their intensity peaks near the 0.87 mm continuum peak.
CO is extended in a north-south direction from the continuum peak, and it traces the protostellar outflows and jets as discussed in \citet{Ikeda2025}. 
The emission distributions of SO, SO$_2$ ($E_u <$100 K),  HCO$^+$, H$^{13}$CN$^+$, and CH$_3$OH ($E_u <$100 K) are extended compared to higher-$E_u$ lines, and these lines have the secondary peak on the south side from the continuum peak (Figure \ref{im_map}).

%We further discuss about the distribution of molecular lines in Section \ref{sec_disc_distri}

\begin{figure*}[tp!]
\begin{center}
\includegraphics[width=18.0cm]{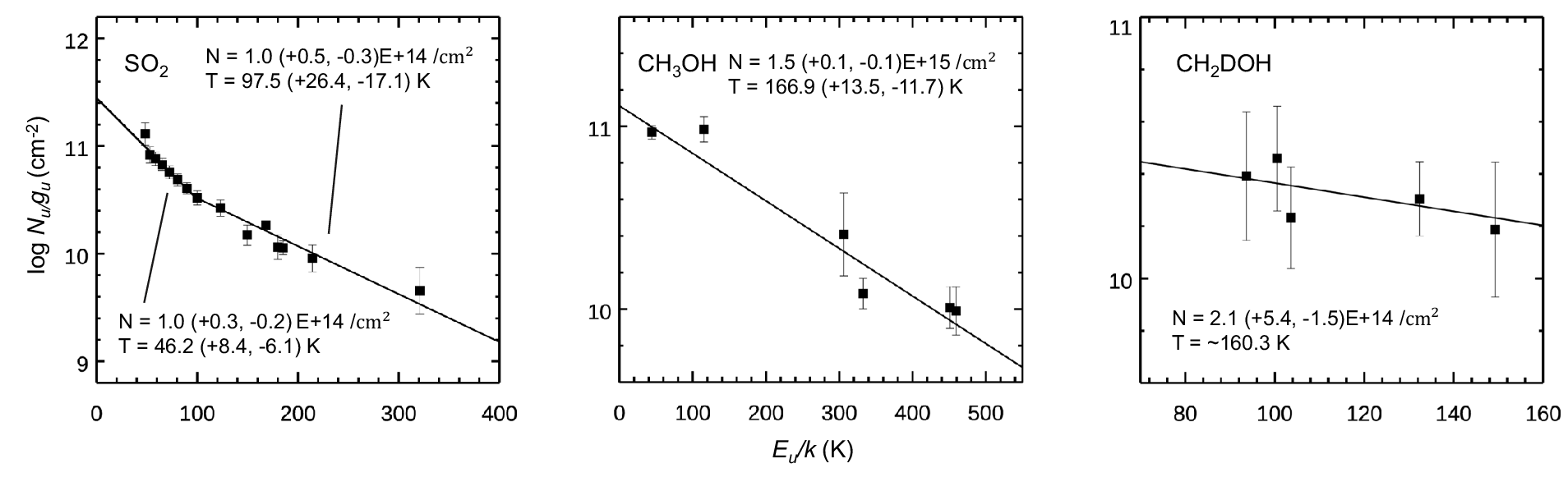}
\caption{The results of the rotational diagram analysis for SO$_2$, CH$_3$OH, and CH$_2$DOH.
The solid lines indicate the results of the linear regression.
The derived column densities and rotational temperatures are shown in each panel.
For SO$_2$, the left line is fitted with using the transitional lines with E$_u$ $<$100 K, while the right solid line is fitted with using the transitional lines with E$_u$ $\geq$100 K.
SO$_2$(13$_{2,12}$--12$_{1,11}$) is removed in the fitting since this line is significantly blended with H$^{13}$CN (4--3).
See Section \ref{sec_rd} for the details.
}
\label{im_RD}
\end{center}
\end{figure*}

\subsection{Column densities and gas temperatures} \label{sec_rd}
Since we detected multiple rotational transitions with different excitation energies for SO$_2$, CH$_3$OH, and CH$_2$DOH, we conducted the rotational diagram analysis to estimate the rotational temperatures and column densities of these molecules.
Here, we assume that the molecular emissions are optically thin and under the condition of the local thermodynamic equilibrium (LTE).
The derived gas density of Sh 2-283-1a SMM1 is $\sim$4 $\times$ 10$^6$ cm$^{-3}$ (Section \ref{sec_h2}), which is a typical value for hot cores \citep[e.g., ][]{van1998, Kurtz2000}. 
\citet[][]{Shimonishi2021} demonstrated that column densities derived from LTE analysis are generally consistent with those from non-LTE analysis within a factor of two for the source with spatial resolution ($\sim$5000 au) and gas density comparable to those in this work.
Thus, we consider the LTE assumption to be reasonable for this source.

%Here, we assume that the observed region is under the condition of the local thermodynamic equilibrium (LTE) and emissions are optically thin.
We use the following formula to perform the rotational diagram analysis in the standard way \citep[e.g.,][]{{Sutton1995, Goldsmith1999}}:
\begin{equation}
\log \left(\frac{ N_{u} }{ g_{u} } \right) = - \left(\frac {\log e}{T_{\mathrm{rot}}} \right) \left(\frac{E_{u}}{k} \right) + \log \left(\frac{N}{Q(T_{\mathrm{rot}})} \right),  \label{Eq_rd1}
\end{equation}
where 
\begin{equation}
\frac{ N_{u} }{ g_{u} } = \frac{ 3 k \int T_{\mathrm{b}} dV }{ 8 \pi^{3} \nu S \mu^{2} }, \label{Eq_rd2} 
\end{equation}
and $N_u$ is the column density of the molecules, $k$ is the Boltzmann constant, $T_\mathrm{b}$ is the brightness temperature, g$_u$ is the degeneracy of the upper state, $T_{\rm{rot}}$ is the rotational temperature, Q($T_{\rm{rot}}$) is the partition function at $T_{\rm{rot}}$, $\nu$ is the rest frequency, $S$ is the line strength, $\mu$ is the dipole moment, and $E_{\rm u}$ is the upper state energy.
All the spectroscopic parameters (Q($T_{\rm{rot}}$), g$_u$, $\nu$, $\mu$, $E_\mathrm{{u}}$, and $S$) are extracted from the CDMS or JPL database.

In the fitting processes, SO$_2$($J = 13_{2,12}$--12$_{1,11}$) was removed from the fitting since this line was significantly blended with H$^{13}$CN($J$ = 4--3).
%For deriving the physical properties of the warm CH$_3$OH component, $J = 4_{1}$ E--3$_{0}$ E ($E_{\rm u}$ = 44 K) was removed in the rotational diagram analysis.
For CH$_2$DOH, we used the a-type ($\Delta$$K_\mathrm{a}$ = 0 and $K_\mathrm{c}$ = 1) R-branch ($\Delta$$J$ = 1) transitions.
While CH$_3$OH and CH$_2$DOH were fitted by a single temperature component, SO$_2$ seemed to have multiple temperature components.
We separated the straight-line fit for SO$_2$ at $E_\mathrm{{u}}$ $<$100 K and $E_\mathrm{{u}}$ $\geq$100 K. 
For the molecular abundance comparisons discussed in Section \ref{sec_abundance_one}, we use the SO$_2$ column density constrained from the lines with $E_\mathrm{{u}}$ $\geq$100 K.
The results of the rotational diagram analysis are shown in Figure \ref{im_RD}.

For molecular species where only a single (or two) lines have been detected and rotational diagram analysis could not be performed, we calculated the column densities by solving Equation \ref{Eq_rd1} for $N$ assuming the rotational temperatures. 
Based on the results of the rotational diagram analysis, we adopted rotational temperatures of 50 K and 150 K for the cold and warm gas components, respectively.
Molecular species with $E_\mathrm{u} \lesssim$100 K and extended emission distributions around the continuum peak were assigned to the cold gas component tracing the relatively low-temperature region associated with a hot core ($T_{\mathrm{rot}}$ = 50 K).
Specifically, we assumed that SO, SO$^+$, NS, HCO$^+$, H$^{13}$CO$^+$, SiO, H$^{13}$CN, HC$^{15}$N, and D$_2$CO trace the cold gas component.
In contrast, molecular species with $E_\mathrm{u} \gtrsim$100 K and compact emission distributions (\textit{trans}-HCOOH, HC$_3$N, H$_2$CCO, $^{13}$CH$_3$OH, CH$_3$SH, and CH$_3$OCH$_3$) were assumed to reside in the warm gas component associated with the hot core ($T_{\mathrm{rot}}$ = 150 K).
The uncertainties of the derived column densities were calculated by combining the fitting errors (corresponding to the 2$\sigma$ level) and the 10$\%$ absolute flux uncertainty in quadrature.
The derived column densities are summarized in Table \ref{tab_N}.

The derived $N$(CH$_3$OH)/$N$($^{13}$CH$_3$OH) ratio for Sh 2-283-1a SMM1 is $\sim$10, which is significantly lower than the local $^{12}$C/$^{13}$C ratio expected from the relation between the galactocentric distance and $^{12}$C/$^{13}$C ratio \citep[$\sim$$80$--$120$ at 15.7 kpc;][]{Yan2023}.
Although the line shapes of CH$_3$OH do not exhibit obvious signs of high optical thickness (see Appendix \ref{sec_appen_fit}), the emission is likely affected by optical thickness, leading to an underestimation of the column density derived from the rotational diagram analysis.
To avoid this underestimation, we estimated $N$(CH$_3$OH) from the $^{13}$CH$_3$OH column density, assuming a $^{12}$C/$^{13}$C ratio of 100 based on the isotope ratio gradients reported by \cite{Milam2005} and \cite{Yan2023} at $D_{\mathrm{GC}}$ = 15.7 kpc.
The resulting $N$(CH$_3$OH) is 1.4 $\times$ 10$^{16}$ cm$^{-2}$, which is approximately one order of magnitude higher than the value obtained from the rotational diagram analysis (Figure \ref{im_RD}).
We use this corrected value of $N$(CH$_3$OH) = 1.4 $\times$ 10$^{16}$ cm$^{-2}$ in the discussion of molecular abundances.
Note that we cannot rule out the possibility that this value is overestimated due to the uncertainties in the assumed $^{12}$C/$^{13}$C ratio and rotational temperature.

%end 

\begin{deluxetable*}{ l c c c c c c c c c c}
\tablecaption{Line Parameters
\label{tab_line}}
\tabletypesize{\footnotesize} %manuscript
%\tabletypesize{\small} %manuscript
\tablehead{
\colhead{Molecule}                       & \colhead{Transition}                                                                                                     &       \colhead{$E_\mathrm{{u}}$} &       \colhead{Frequency} &        \colhead{$T_\mathrm{{b}}$} &     \colhead{$\Delta$$V$} &     \colhead{$\int T_\mathrm{{b}} dV$} &       \colhead{$V_\mathrm{{LSR}}$} &        \colhead{RMS} &       \colhead{Note} \\
\colhead{ }                              & \colhead{ }                                                                                                              &        \colhead{(K)} &           \colhead{(GHz)} &             \colhead{(K)} &          \colhead{(km s$^{-1}$)} &             \colhead{(K km s$^{-1}$)} &          \colhead{(km s$^{-1}$)} &        \colhead{(K)} &           \colhead{}
}
\startdata
 CO                                      &  3--2                                                                                                                    &   33       & 345.79599       &  $\sim$16  &    \nodata          & $\sim$157    &  \nodata            & 0.02 &    \nodata \\
 SO                                      &  $N_J$ = 8$_{8}$--7$_{7}$                                                                                                &   87       & 344.31061       &   2.82 $\pm$   0.04  &    2.6          & 7.88 $\pm$ 0.12      &  53.0           & 0.02 &    \nodata \\
 SO                                      &  $N_J$ = 8$_{9}$--7$_{8}$                                                                                                &   79       & 346.52848       &   4.26 $\pm$   0.04  &    2.5          & 11.41 $\pm$ 0.12     &  53.1           & 0.02 &    \nodata \\
 SO$^+$                                  &  8$_{-1,8}$--7$_{1,7}$                                                                                                   &   70       & 347.74001       &   0.13 $\pm$   0.05  &    1.6          & 0.21 $\pm$ 0.08      &  52.9           & 0.02 &    \nodata \\
 SO$_2$                                  &  4$_{4,0}$--4$_{3,1}$                                                                                                    &   48       & 358.03789       &   0.22 $\pm$   0.04  &    2.8          & 0.68 $\pm$ 0.16      &  53.0           & 0.03 &    \nodata \\
 SO$_2$                                  &  5$_{4,2}$--5$_{3,3}$                                                                                                    &   53       & 358.01315       &   0.32 $\pm$   0.05  &    2.4          & 0.80 $\pm$ 0.14      &  53.2           & 0.03 &    \nodata \\
 SO$_2$                                  &  6$_{4,2}$--6$_{3,3}$                                                                                                    &   59       & 357.92585       &   0.34 $\pm$   0.04  &    2.8          & 1.04 $\pm$ 0.15      &  53.4           & 0.03 &    \nodata \\
 SO$_2$                                  &  7$_{4,4}$--7$_{3,5}$                                                                                                    &   65       & 357.89244       &   0.35 $\pm$   0.04  &    3.1          & 1.16 $\pm$ 0.15      &  53.3           & 0.03 &    \nodata \\
 SO$_2$                                  &  8$_{4,4}$--8$_{3,5}$                                                                                                    &   72       & 357.58145       &   0.30 $\pm$   0.04  &    3.7          & 1.19 $\pm$ 0.17      &  53.4           & 0.03 &    \nodata \\
 SO$_2$                                  &  9$_{4,6}$--9$_{3,7}$                                                                                                    &   81       & 357.67182       &   0.29 $\pm$   0.04  &    3.8          & 1.18 $\pm$ 0.16      &  53.5           & 0.03 &    \nodata \\
 SO$_2$                                  &  10$_{4,6}$--10$_{3,7}$                                                                                                  &   90       & 356.75519       &   0.31 $\pm$   0.04  &    3.5          & 1.13 $\pm$ 0.14      &  53.2           & 0.03 &    (1) \\
 SO$_2$                                  &  11$_{4,8}$--11$_{3,9}$                                                                                                  &  100       & 357.38758       &   0.25 $\pm$   0.04  &    3.9          & 1.04 $\pm$ 0.16      &  53.6           & 0.03 &    \nodata \\
 SO$_2$                                  &  13$_{2,12}$--12$_{1,11}$                                                                                                &   93       & 345.33854       &   0.56 $\pm$   0.03  &    3.7          & 2.19 $\pm$ 0.12      &  52.9           & 0.02 &    (2) \\
 SO$_2$                                  &  13$_{4,10}$--13$_{3,11}$                                                                                                &  123       & 357.16539       &   0.20 $\pm$   0.04  &    4.8          & 1.01 $\pm$ 0.18      &  53.7           & 0.03 &    \nodata \\
 SO$_2$                                  &  15$_{4,12}$--15$_{3,13}$                                                                                                &  150       & 357.24119       &   0.16 $\pm$   0.04  &    3.9          & 0.67 $\pm$ 0.15      &  53.5           & 0.03 &    \nodata \\
 SO$_2$                                  &  17$_{4,14}$--17$_{3,15}$                                                                                                &  180       & 357.96290       &   0.17 $\pm$   0.04  &    3.3          & 0.60 $\pm$ 0.16      &  53.9           & 0.03 &    \nodata \\
 SO$_2$                                  &  19$_{1,19}$--18$_{0,18}$                                                                                                &  168       & 346.65217       &   0.35 $\pm$   0.03  &    4.3          & 1.61 $\pm$ 0.14      &  53.4           & 0.02 &    \nodata \\
 SO$_2$                                  &  19$_{4,16}$--19$_{3,17}$                                                                                                &  214       & 359.77068       &   0.16 $\pm$   0.04  &    3.2          & 0.54 $\pm$ 0.15      &  53.8           & 0.03 &    \nodata \\
 SO$_2$                                  &  20$_{0,20}$--19$_{1,19}$                                                                                                &  185       & 358.21563       &   0.25 $\pm$   0.04  &    4.1          & 1.09 $\pm$ 0.16      &  53.5           & 0.03 &    \nodata \\
 SO$_2$                                  &  25$_{3,23}$--25$_{2,24}$                                                                                                &  321       & 359.15116       &   0.09 $\pm$   0.04  &    3.1          & 0.28 $\pm$ 0.14      &  53.5           & 0.03 &    \nodata \\
 NS                                      &  8$_{1,8,9}$--7$_{-1,7,8}$                                                                                               &   70       & 346.22014       &   0.05 $\pm$   0.03  &    3.5          & 0.20 $\pm$ 0.10      &  52.9           & 0.02 &    $^{\mathrm{(a)}}$(3) \\
 %HDCS                                    &  11$_{1,10}$--10$_{1,9}$                                                                                                 &  109       & 347.30457       &   0.12 $\pm$   0.02  &   12.8          & 1.60 $\pm$ 0.20      &  54.9           & 0.02 &    \nodata \\
 SiO                                     &  8--7                                                                                                                    &   75       & 347.33058       &   0.38 $\pm$   0.03  &    5.9          & 2.36 $\pm$ 0.15      &  53.3           & 0.02 &    \nodata \\
 HCO$^+$                                 &  4--3                                                                                                                    &   43       & 356.73422       &  18.30 $\pm$   0.05  &    2.7          & 52.99 $\pm$ 0.20     &  53.0           & 0.03 &    \nodata \\
 H$^{13}$CO$^+$                          &  4--3                                                                                                                    &   42       & 346.99834       &   1.49 $\pm$   0.04  &    2.1          & 3.26 $\pm$ 0.13      &  53.1           & 0.02 &    \nodata \\
 H$^{13}$CN                              &  4--3                                                                                                                    &   41       & 345.33977       &   0.56 $\pm$   0.03  &    3.7          & 2.19 $\pm$ 0.12      &  54.0           & 0.02 &    \nodata \\
 HC$^{15}$N                              &  4--3                                                                                                                    &   41       & 344.20011       &   0.20 $\pm$   0.03  &    3.2          & 0.69 $\pm$ 0.12      &  53.5           & 0.02 &    \nodata \\
 D$_2$CO                                 &  6$_{2,4}$--5$_{2,3}$                                                                                                    &   81       & 357.87145       &   0.12 $\pm$   0.05  &    2.5          & 0.32 $\pm$ 0.14      &  54.2           & 0.03 &    \nodata \\
 \textit{trans}-HCOOH                    &  16$_{0,16}$--15$_{0,15}$                                                                                                &  143       & 345.03056       &   0.10 $\pm$   0.03  &    4.0          & 0.43 $\pm$ 0.12      &  53.2           & 0.02 &    \nodata \\
 \textit{trans}-HCOOH                    &  15$_{2,13}$--14$_{2,12}$                                                                                                &  144       & 346.71886       &   0.09 $\pm$   0.03  &    3.5          & 0.35 $\pm$ 0.12      &  54.3           & 0.02 &    \nodata \\
% \textit{cis}-HCOOH                      &  16$_{3,13}$--16$_{2,14}$                                                                                                &  176       & 345.03248       &   0.10 $\pm$   0.04  &    3.0          & 0.33 $\pm$ 0.10      &  54.4           & 0.02 &    \nodata \\
 HC$_3$N                                 &  38--37                                                                                                                  &  323       & 345.60901       &   0.05 $\pm$   0.03  &    4.2          & 0.24 $\pm$ 0.15      &  53.1           & 0.02 &    $^{\mathrm{(a)}}$ $^{\mathrm{(b)}}$ \\
 H$_2$CCO                                &  17$_{1,16}$--16$_{1,15}$                                                                                                &  163       & 346.60045       &   0.05 $\pm$   0.04  &    2.5          & 0.14 $\pm$ 0.10      &  54.3           & 0.02 &    $^{\mathrm{(a)}}$ \\
 CH$_3$OH                                &  4$_{1}$ E--3$_{0}$ E                                                                                                    &   44       & 358.60580       &   0.55 $\pm$   0.04  &    3.0          & 1.76 $\pm$ 0.15      &  53.1           & 0.03 &    \nodata \\
 CH$_3$OH                                &  5$_{4}$ A$^-$--6$_{3}$ A$^-$                                                                                            &  115       & 346.20272       &   0.17 $\pm$   0.03  &    4.2          & 0.79 $\pm$ 0.13      &  53.2           & 0.02 &    (4) \\
 CH$_3$OH                                &  10$_{6}$ E--11$_{5}$ E                                                                                                  &  306       & 358.41465       &   0.08 $\pm$   0.04  &    3.5          & 0.30 $\pm$ 0.15      &  54.0           & 0.03 &    $^{\mathrm{(c)}}$ \\
 CH$_3$OH                                &  16$_{1}$ A$^-$--15$_{2}$ A$^-$                                                                                          &  333       & 345.90392       &   0.14 $\pm$   0.03  &    4.9          & 0.72 $\pm$ 0.14      &  54.0           & 0.02 &    $^{\mathrm{(c)}}$ \\
 CH$_3$OH                                &  18$_{-3}$ E--17$_{-4}$ E                                                                                                &  459       & 345.91926       &   0.07 $\pm$   0.03  &    6.0          & 0.45 $\pm$ 0.14      &  53.5           & 0.02 &    $^{\mathrm{(c)}}$ \\
 CH$_3$OH                                &  19$_{1}$ A$^+$--18$_{2}$ A$^+$                                                                                          &  451       & 344.44343       &   0.11 $\pm$   0.03  &    4.2          & 0.50 $\pm$ 0.13      &  53.5           & 0.02 &    $^{\mathrm{(c)}}$ \\
 $^{13}$CH$_3$OH                         &  7$_{2}$ E--6$_{1}$ E                                                                                                    &   86       & 357.65795       &   0.13 $\pm$   0.05  &    2.2          & 0.30 $\pm$ 0.12      &  53.2           & 0.03 &    \nodata \\
 CH$_2$DOH                               &  2$_{1,1}$$e_1$--2$_{0,2}$$e_0$                                                                                          &   23       & 344.86895       &   0.06 $\pm$   0.04  &    2.8          & 0.17 $\pm$ 0.10      &  53.7           & 0.02 &    $^{\mathrm{(a)}}$ $^{\mathrm{(c)}}$ \\
 CH$_2$DOH                               &  3$_{1,2}$$e_1$--3$_{0,3}$$e_0$                                                                                          &   29       & 346.25650       &   0.05 $\pm$   0.03  &    3.8          & 0.21 $\pm$ 0.16      &  54.1           & 0.02 &    $^{\mathrm{(a)}}$ $^{\mathrm{(c)}}$\\
 CH$_2$DOH                               &  7$_{1,6}$$e_1$--7$_{0,7}$$e_0$                                                                                          &   77       & 357.52856       &   0.09 $\pm$   0.05  &    2.1          & 0.21 $\pm$ 0.13      &  54.1           & 0.03 &    $^{\mathrm{(c)}}$ \\
 CH$_2$DOH                               &  8$_{0,8}$$o_1$--7$_{0,7}$$o_1$                                                                                          &   95       & 356.98158       &   0.10 $\pm$   0.05  &    3.2          & 0.34 $\pm$ 0.13      &  55.2           & 0.03 &    (5) \\
 CH$_2$DOH                               &  8$_{1,7}$$e_2$--7$_{1,6}$$e_2$                                                                                          &  101       & 359.75351       &   0.06 $\pm$   0.03  &    5.7          & 0.38 $\pm$ 0.17      &  55.0           & 0.02 &    $^{\mathrm{(b)}}$ \\
 CH$_2$DOH                               &  8$_{2,6}$$e_0$--7$_{2,5}$$e_0$                                                                                          &   94       & 359.72880       &   0.10 $\pm$   0.05  &    2.3          & 0.25 $\pm$ 0.14      &  53.2           & 0.03 &    \nodata \\
 CH$_2$DOH                               &  8$_{2,7}$$e_1$--7$_{2,6}$$e_1$                                                                                          &  104       & 356.89967       &   0.07 $\pm$   0.03  &    5.9          & 0.41 $\pm$ 0.18      &  51.5           & 0.02 &   $^{\mathrm{(b)}}$ (6) \\
 CH$_2$DOH                               &  8$_{3,6}$$e_2$--7$_{3,5}$$e_2$                                                                                          &  132       & 357.02078       &   0.11 $\pm$   0.04  &    4.0          & 0.46 $\pm$ 0.15      &  52.4           & 0.03 &    (7) \\
  CH$_2$DOH                               &  8$_{4,4}$$e_1$--7$_{4,3}$$e_1$                                                                                          &  149       & 356.93241       &  0.07 $\pm$   0.04  &    3.7          & 0.30 $\pm$ 0.17      &  53.5           & 0.03 &   $^{\mathrm{(b)}}$ (8) \\
  CH$_3$SH                                &  14$_{1,13,1}$--13$_{1,12,1}$                                                                                            &  134       & 356.62701       &   0.10 $\pm$   0.05  &    2.7          & 0.28 $\pm$ 0.15      &  53.8           & 0.03 &    $^{\mathrm{(a)}}$ \\
 CH$_3$OCH$_3$                           &  5$_{5,0}$--4$_{4,0}$ EE                                                                                                 &   49       & 358.45202       &   0.13 $\pm$   0.04  &    3.9          & 0.52 $\pm$ 0.14      &  52.0           & 0.03 &    (9) \\
 CH$_3$OCH$_3$                           &  12$_{3,10}$--11$_{2,9}$ EE                                                                                              &   84       & 359.38459       &   0.08 $\pm$   0.03  &    4.4          & 0.39 $\pm$ 0.15      &  54.1           & 0.02 &  $^{\mathrm{(b)}}$  (10) \\
\enddata

\tablecomments{The uncertainties of the line parameters are at the 2$\sigma$ level derived from Gaussian fitting.
$^{\mathrm{(a)}}$Tentative detection. 
$^{\mathrm{(b)}}$Two channels were binned to reduce noise.
$^{\mathrm{(c)}}$The Savitzky-Golay filter was applied to reduce noise.
(1) Partially blend with HCO$^+$  (4--3). 
(2) Blend with SO$_2$ (13$_{2,12}$--12$_{1,11}$).
(3) Blended with three hyperfine components. 
(4) Blended with two hyperfine components. 
(5) Blended with five hyperfine components. 
(6) Blended with three hyperfine components. 
(7) Blended with two hyperfine components. 
(8) Blended with two hyperfine components. 
(9) Blended with six hyperfine components. 
(10) Blended with four hyperfine components. 
}
\end{deluxetable*}

\subsection{Column density of molecular hydrogen} \label{sec_h2}
The column density of molecular hydrogen ($N_{\rm{H_2}}$) is estimated using the dust continuum data.
We used the following formula based on the standard treatment of optically thin dust emission:
\begin{equation}
N_{\mathrm{H_2}} = \frac{F_{\nu} / \Omega}{2 \kappa_{\nu} B_{\nu}(T_{\rm{d}}) Z \mu m_{\mathrm{H}}} \label{Eq_h2}, 
\end{equation}
and $F_{\nu}/\Omega$ is the continuum flux density per beam solid angle as estimated from the observations, and $\kappa_{\nu}$ is the mass absorption coefficient of dust grains coated by thin ice mantles taken from \citet{Oss94} and we use 1.89 cm$^2$ g$^{-1}$ at 870 $\mu$m.
$T_\mathrm{d}$ is the dust temperature, $B_{\nu}(T_{\rm{d}})$ is the Planck function at $T = T_{\rm{d}}$, $Z$ is the dust-to-gas mass ratio, $\mu$ is the mean atomic mass per hydrogen \citep[1.41;][]{Cox00}, and $m_{\mathrm{H}}$ is the hydrogen mass. 
We used the dust-to-gas mass ratio of 0.0024, derived by scaling the canonical local ISM value of 0.008 with the metallicity of the target ($\sim$0.3 $Z_\odot$).
The uncertainty in the dust temperature introduces uncertainty in the derivation of $N_{\rm{H_2}}$.
We assumed the dust temperature to be equal to that of the high-temperature molecular gas component (150 K), based on the results of the rotational diagram analysis (Section \ref{sec_rd}).
%The derived $N_{\rm{H_2}}$ is 1.4 $\times$ 10$^{23}$ cm$^{-2}$ (ranging from (0.68--4.6) $\times$ 10$^{23}$ cm$^{-2}$ for dust temperatures of 50--300 K).
The derived $N_{\rm{H_2}}$ is 1.4 $\times$ 10$^{23}$ cm$^{-2}$, with a range of (1.1-2.7) $\times$ 10$^{23}$ cm$^{-2}$ for dust temperatures of 80--180 K based on the $T_\mathrm{rot}$ of SO$_2$ and CH$_3$OH (Figure \ref{im_RD}).
Assuming non-ice dust, the derived $N_{\rm{H_2}}$ is 7.4 $\times$ 10$^{22}$ cm$^{-2}$ with $\kappa_{\nu}$ = 3.54 cm$^2$ g$^{-1}$ at 870 $\mu$m and $T_\mathrm{d}$ = 150 K.
In this paper, we adopt $N_{\rm{H_2}}$ = 1.4$^{+ 1.3}_{-0.3}$ $\times$ 10$^{23}$ cm$^{-2}$ as the representative value of H$_2$ column density.
The gas density of the core was estimated by dividing the $N_{\rm{H_2}}$ by its extent (6.5 $\times$ 10$^3$ au is assumed), and the derived value is 3.7 $\times$ 10$^6$ cm$^{-3}$.

After determining $N_{\rm{H_2}}$, we derived the fractional abundances of the detected molecular species.
The derived molecular abundances are summarized in Table \ref{tab_N}.
The uncertainties for the abundance ratios used in Section \ref{sec_disc} were calculated using standard error propagation rules.

\begin{deluxetable}{ l c c c c}
\tablecaption{Estimated Rotational Temperatures, Column Densities, and Fractional Abundances 
\label{tab_N}}
\tabletypesize{\footnotesize} %manuscript
%\tabletypesize{\small} %manuscript
\tablehead{
\colhead{Molecule}   &             \colhead{$T_\mathrm{{rot}}$}       &       \colhead{$N$(X)} &              \colhead{$N$(X)/$N$(H$_2$)} \\
\colhead{ }          &           \colhead{(K)}       &            \colhead{(cm$^{-2}$)} &                              \colhead{ }
}
\startdata
H$_2$      &   --     & (1.4 $^{+ 1.3}_{-0.3}$) $\times$ 10$^{23}$$^{\mathrm{(a)}}$      & --               \\
 SO             &  50   & (1.5 $\pm$ 0.2) $\times$ 10$^{14}$      & 1.1   $\times$ 10$^{-10}$                 \\
   SO$^+$       &   50    & (5.2 $\pm$ 2.0) $\times$ 10$^{12}$      & 3.8   $\times$ 10$^{-11}$                 \\
   SO$_2$        &  46.2$^{+ 8.4}_{- 6.1}$    & (1.0 $^{+ 0.3}_{-0.2}$) $\times$ 10$^{14}$      & 7.3   $\times$ 10$^{-10}$                 \\
                       &  97.5$^{+ 26.4}_{- 17.1}$    & (1.0 $^{+ 0.5}_{-0.3}$) $\times$ 10$^{14}$      & 7.3  $\times$ 10$^{-10}$                 \\
  NS           &   50    & (1.5 $\pm$ 0.8) $\times$ 10$^{12}$      & 1.1   $\times$ 10$^{-11}$                 \\
% HDCS          &   150   & (7.5 $\pm$ 0.9) $\times$ 10$^{13}$      & 2.0   $\times$ 10$^{-10}$                 \\
 HCO$^+$       &   50   & (2.3 $\pm$ 0.2) $\times$ 10$^{13}$      & 1.7   $\times$ 10$^{-10}$                 \\
 H$^{13}$CO$^+$  &  50  & (1.5 $\pm$ 0.2) $\times$ 10$^{12}$      & 1.1  $\times$ 10$^{-11}$                 \\
 SiO          &  50     & (3.2 $\pm$ 0.4) $\times$ 10$^{12}$      & 2.3   $\times$ 10$^{-11}$                 \\
 H$^{13}$CN     &  50   & (1.7 $\pm$ 0.2) $\times$ 10$^{12}$      & 1.2 $\times$ 10$^{-11}$                 \\
 HC$^{15}$N   &   50    & (5.3 $\pm$ 1.0) $\times$ 10$^{11}$      & 3.9  $\times$ 10$^{-12}$                 \\
D$_2$CO      &   50    & (3.4 $\pm$ 1.6) $\times$ 10$^{12}$      & 2.5 $\times$ 10$^{-11}$                 \\
 \textit{trans}-HCOOH     &   150   & (4.7 $\pm$ 1.7) $\times$ 10$^{13}$      & 3.4   $\times$ 10$^{-10}$                 \\
% \textit{cis}-HCOOH           &    150   & (2.3 $\pm$ 0.7) $\times$ 10$^{13}$      & 6.2   $\times$ 10$^{-11}$                 \\
HC$_3$N      &  150     & (1.3 $\pm$ 0.8) $\times$ 10$^{12}$      & 9.6  $\times$ 10$^{-12}$                 \\
H$_2$CCO      &  150    & (7.3 $\pm$ 5.4) $\times$ 10$^{12}$      & 5.3  $\times$ 10$^{-11}$                 \\
CH$_3$OH     &   166.9$^{+ 13.5}_{- 11.7}$    & (1.5 $^{+ 0.1}_{-0.1}$) $\times$ 10$^{15}$      & 1.1   $\times$ 10$^{-8}$                  \\
              &    150$^{\mathrm{(b)}}$    & (1.4 $\pm$ 0.6) $\times$ 10$^{16}$      & 1.0   $\times$ 10$^{-7}$                 \\
$^{13}$CH$_3$OH   &  150    & (1.4 $\pm$ 0.6) $\times$ 10$^{14}$      & 1.0   $\times$ 10$^{-9}$                 \\
 CH$_2$DOH      & $\sim$160 %160.3$^{+ 391.4}_{- 100.8}$ 
 & (2.1$^{+ 5.4}_{-1.5}$) $\times$ 10$^{14}$      & 1.5   $\times$ 10$^{-9}$                 \\
CH$_3$SH      &   150   & (7.9 $\pm$ 4.4) $\times$ 10$^{13}$      & 5.8  $\times$ 10$^{-10}$                 \\
CH$_3$OCH$_3$    &   150   & (2.0 $\pm$ 0.6) $\times$ 10$^{14}$      & 1.5   $\times$ 10$^{-9}$                 \\
\enddata
\tablecomments{The uncertainties of the column densities were calculated by combining the fitting errors (corresponding to the 2$\sigma$ level) and the 10$\%$ absolute flux uncertainty in quadrature.
$^{\mathrm{(a)}}$The uncertainty corresponds to the range of $N$(H$_2$) derived assuming the dust temperatures of 80–180 K, consistent with the rotational temperatures of SO$_2$ and CH$_3$OH.
$^{\mathrm{(b)}}$The column density of CH$_3$OH used for the abundance comparisons derived from that of $^{13}$CH$_3$OH by scaling the $^{12}$C/$^{13}$C ratio (assuming $^{12}$C/$^{13}$C = 100).}
\end{deluxetable}

\begin{figure}[tp!]
\begin{center}
\includegraphics[width=8.0cm]{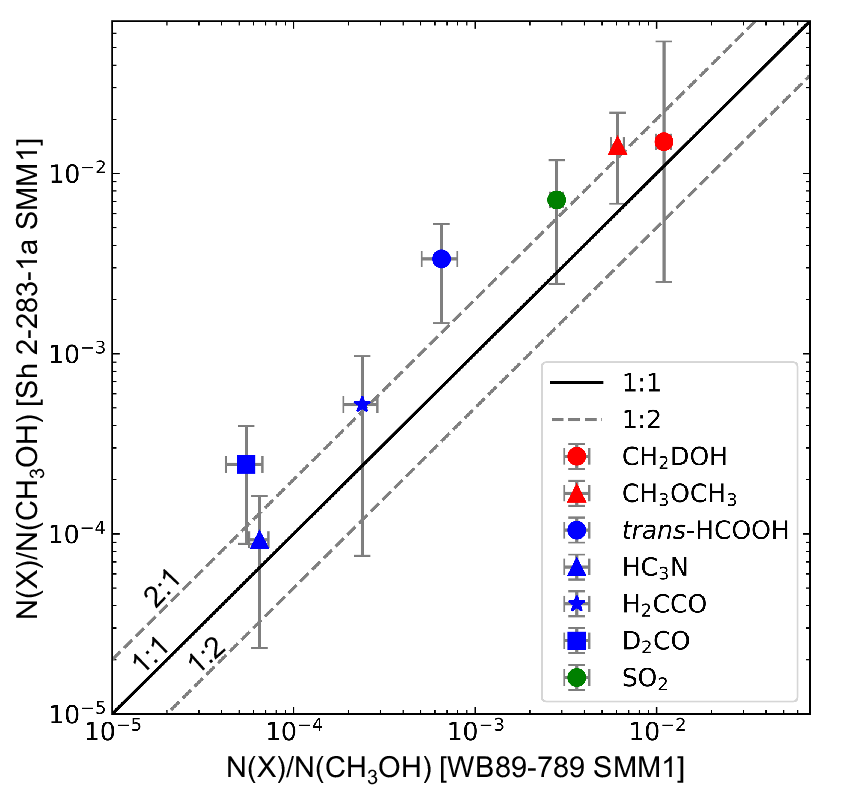}
\caption{Comparison of the molecular abundances normalized by CH$_3$OH column density between the outer Galactic hot cores.
%and (b) N113 A1 (LMC) with Sh 2-283-1a SMM1.
%The molecular species used for the comparison are shown in the upper left in each panel.
The dotted lines represent the abundance ratio of 2:1 and 1:2 for Sh 2-283-1a SMM1, while the solid line represents that of 1:1.
The molecular abundances of WB89-789 SMM1 are taken from \cite{Shimonishi2021}.
%The dotted lines in  panel (b) represent the abundance ratio of 100:1, 10:1, 1:10, and 1:100 for Sh 2-283-1a SMM1, while the solid line represents that of 1:1.
}
\label{im_Abun}
\end{center}
\end{figure}

\begin{figure*}[tp!]
\begin{center}
\includegraphics[width=18.0cm]{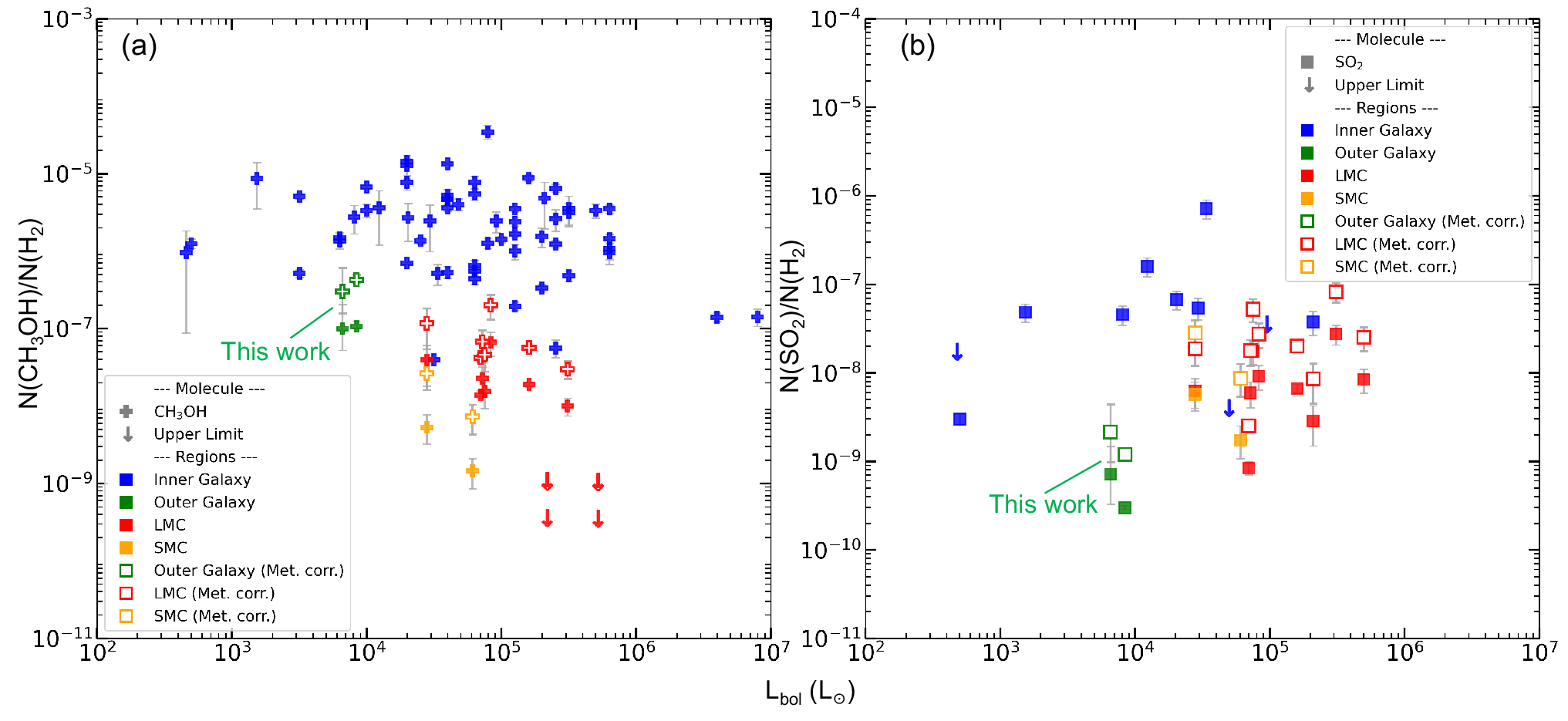}
\caption{The $N$(CH$_3$OH)/$N$(H$_2$) (a) and $N$(SO$_2$)/$N$(H$_2$) (b) ratios as a function of luminosities for the inner Galactic sources \citep[blue;][]{Fuente2014,Qin2022,Gelder2022L,Santos2024,Chen2025ATMs,Li2025}, the outer Galactic sources \citep[green;][]{Shimonishi2021}, the LMC sources \citep[red;][]{Shimonishi2016, Shimonishi2020,Shimonishi2021,Shimonishi_magos,Sewilo2022, Golshan2024, Broadmeadow2025}, and the SMC sources \citep[orange;][]{Shimonishi2023} .
The open marks in white show the values corrected for metallicity for each low-metallicity source (e.g., multiplied by a factor of 3 for $Z = 1/3 $ $Z_\odot$).
The subscript arrows in each plot represent the upper limits.
}
\label{im_Abundance_met_corr}
\end{center}
\end{figure*}

\begin{figure}[tp!]
\begin{center}
\includegraphics[width=8.0cm]{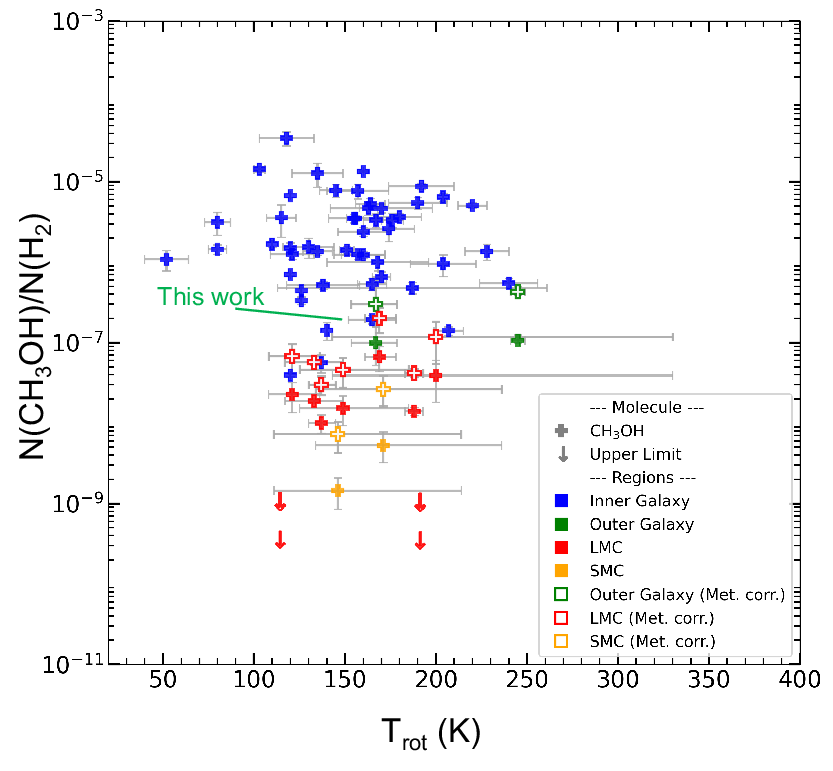}
\caption{$N$(CH$_3$OH)/$N$(H$_2$) ratios as a function of rotational temperatures for the inner Galactic hot cores \citep[blue;][]{Fuente2014,Li2025,Chen2025ATMs}, outer Galactic hot cores \citep[green;][]{Shimonishi2021}, the LMC hot cores \citep[red;][]{Shimonishi2016, Shimonishi2020,Shimonishi_magos, Broadmeadow2025}, and SMC hot cores \citep[orange;][]{Shimonishi2023}.
%Shimonishi+2025を追加
The rotational temperatures are derived from the CH$_3$OH lines.
For the inner Galactic sources whose rotational temperatures of CH$_3$OH did not be derived, rotational temperatures of the other COMs are used.
For the LMC and SMC sources whose rotational temperatures of CH$_3$OH could not be derived, rotational temperatures of SO$_2$ are used.
The open marks in white show the values corrected for metallicity for each low-metallicity hot core.
The subscript arrows represent the upper limits.
}
\label{im_Abun_Trot}
\end{center}
\end{figure}

\section{Discussion} \label{sec_disc} 
\subsection{Hot molecular core associated with Sh 2-283-1a SMM1} \label{sec_disc_star} 
Sh 2-283-1a SMM1 exhibits (1) a compact distribution of envelope gas ($\sim$0.03 pc, Section \ref{sec_spc}), (2) a high gas temperature ($>$100 K, Section \ref{sec_rd}), (3) a high gas density ($\sim$4 ${\times}$ 10$^{6}$ cm$^{-3}$, Section \ref{sec_h2}), and (4) the detection of various molecular species, including COMs.
These characteristics suggest that this source is associated with a hot core.
The luminosity of Sh 2-283-1a SMM1 is estimated to be 10$^3$--10$^4$ L$_\sun$ \citep{Ikeda2025}, suggesting that it hosts an intermediate- or high-mass protostar.

\subsection{Chemical properties of the outer Galactic hot cores} \label{sec_abundance_one} 

Figure \ref{im_Abun} compares the molecular abundances normalized by $N$(CH$_3$OH) between Sh 2-283-1a SMM1 and WB89-789 SMM1. 
The latter is the only previously reported hot core in the outer Galaxy \citep[$D_{\rm{GC}}$ = 19.0 kpc;][]{Shimonishi2021}.
%and (b) N105 2A (LMC) with those of Sh 2-283-1a SMM1.
%N105 2A is the most chemically rich hot core presently detected in the LMC, and even COMs larger than CH$_3$OH (e.g., CH$_3$OCH$_3$) were detected \citep{Sewilo2022}.
Since CH$_3$OH lines with $E_{\rm u}$ $\geq$100 K usually trace the hot gas arising from a high-temperature region where products of grain surface chemistry (e.g., COMs) are sublimated from dust mantles, such a comparison is useful for evaluating the chemical properties of hot cores \citep{Herbst2009}.
%especially for organic molecules \citep{Herbst2009}.
Moreover, we can mitigate the effect of metallicity differences by normalizing with CH$_3$OH. 

In this figure, we compare the molecular abundances for 7 molecular species detected in both sources: two COMs (CH$_2$DOH and CH$_3$OCH$_3$; red), four organic molecules (\textit{trans}-HCOOH, HC$_3$N, H$_2$CCO, and D$_2$CO; blue), and SO$_2$ (green).
Note that HC$_3$N and H$_2$CCO are tentatively detected in Sh 2-283-1a SMM1. 
Although the molecular abundances normalized by CH$_3$OH are overall higher in Sh 2-283-1a SMM1 by roughly a factor of two, these two outer Galactic hot cores share a similar chemical composition. 
This indicates that Sh 2-283-1a SMM1 exhibits chemical complexity comparable to that of WB89-789 SMM1 reported by \cite{Shimonishi2021}, which is located in a different region of the outer Galaxy.

%These results indicate that the chemical properties of hot cores in low-metallicity environments would not simply be scaled with their metallicity, and additional factors affect them, as argued in the previous studies \citep{Shimonishi2020, Shimonishi2021}.

\subsection{Molecular abundances: Comparison with the Galactic and Magellanic hot cores} \label{sec_abundance_two} 
Figure \ref{im_Abundance_met_corr} compares the molecular abundances of hot cores (CH$_3$OH and SO$_2$) relative to H$_2$ as a function of luminosity.
%Blue symbols represent the inner Galactic hot cores, while green, red, and orange symbols denote the outer Galactic, LMC, and SMC hot cores, respectively.
%White open symbols show the metallicity-corrected values for each low-metallicity source (e.g., multiplied by a factor of 3 for $Z = 1/3 $ $Z_\odot$).
The CH$_3$OH and SO$_2$ column densities for the inner Galactic sources were taken from \cite{Fuente2014,Gelder2022L,Qin2022,Li2025} and \cite{Fuente2014,Santos2024}.
These sources have $D_{\rm{GC}}$ ranging from $\sim$0.2--8 kpc (Appendix \ref{sec_appen_abun}).
%Since \cite{Santos2024} did not derive H$_2$ column densities, we calculated them from the continuum data summarized in \cite{Gelder2022L}, following the method in Section \ref{sec_h2}.
We calculated the H$_2$ column densities for the sources in \cite{Santos2024} by using the continuum data summarized in \cite{Gelder2022L}, following the method used in Section \ref{sec_h2}.
We assumed the dust temperature is equal to the CH$_3$OH excitation temperature reported in \cite{Gelder2022}.

The column densities of CH$_3$OH and SO$_2$ for the outer Galactic and Magellanic sources are obtained from the literature \citep[][]{Shimonishi2016, Shimonishi2020,Shimonishi2021,Shimonishi2023, Shimonishi_magos,Sewilo2022, Golshan2024, Broadmeadow2025}.
%The H$_2$ column densities are re-calculated for the sources whose H$_2$ column densities are derived assuming low temperature ($T_\mathrm{{dust}}$ $<$ 100 K) in the literature, and the assumed dust temperatures are the same as the high-temperature molecular gas derived from the rotational analyses. 
For some sources where H$_2$ column densities were originally derived assuming $T_\mathrm{dust} <$100 K, we recalculated them assuming that the dust temperature is equal to the rotational temperature of either CH$_3$OH or SO$_2$.

Figure \ref{im_Abundance_met_corr}(a) shows a comparison of the $N$(CH$_3$OH)/$N$(H$_2$) ratios among the sources described above.
This figure demonstrates that the $N$(CH$_3$OH)/$N$(H$_2$) ratios vary significantly among the sources, and the abundances exhibit a large scatter, spanning approximately two orders of magnitude, even when limited to the inner Galactic sources.
No correlation is found between the $N$(CH$_3$OH)/$N$(H$_2$) ratios and luminosity.
The $N$(CH$_3$OH)/$N$(H$_2$) ratios of the outer Galactic sources are mutually similar, and the values can be approximately scaled with their metallicities when compared to the inner Galactic sources ($\sim$1/4 of the median of the inner Galactic sources).
However, for the Magellanic sources, the values are lower or, at best, lie near the lower limit of the inner Galactic sources, even when the values are corrected for their metallicities; the median values of abundances are only $\sim$1/27 (LMC) and $\sim$1/90 (SMC) of that of the inner Galactic sources.

Figure \ref{im_Abun_Trot} shows the $N$(CH$_3$OH)/$N$(H$_2$) ratios as a function of $T_{\rm{rot}}$ of CH$_3$OH.
In this figure, for the inner Galactic sources for which rotational temperatures of CH$_3$OH were not derived, the rotational temperatures of other COMs are used.
For the LMC and SMC sources for which rotational temperatures of CH$_3$OH were not derived, the rotational temperatures of SO$_2$ are used.
This figure indicates no correlation between the $N$(CH$_3$OH)/$N$(H$_2$) ratios and gas temperatures, suggesting that the $N$(CH$_3$OH)/$N$(H$_2$) ratio is independent of the current physical conditions of the sources.
It is widely accepted that CH$_3$OH in hot cores is mainly formed on dust grain surfaces during the initial ice-forming stages via successive hydrogenation of CO, and as star formation proceeds, it is released into the gas phase via thermal desorption \citep[e.g., ][]{Wat02}.
%The formation efficiency of CH$_3$OH is heavily affected by the environment of these initial ice-forming stages, especially regarding dust temperature.
%Since the dust temperature of $\lesssim$ 15 K is needed for the CO depletion to the dust grain surfaces and to keep hydrogen atoms in the solid phase, such a low temperature environment is essential for the efficient formation of CH$_3$OH in protostellar regions.
The formation efficiency of CH$_3$OH is significantly affected by the surrounding environment during these initial ice-forming stages, particularly by dust temperature.
Since dust temperatures of $\lesssim$20 K are required for CO depletion onto dust grain surfaces and to maintain hydrogen atoms in the solid phase, such a low-temperature environment is essential for the efficient formation of CH$_3$OH in protostellar regions.
%These points suggest that the outer Galactic hot cores likely experienced environments similar to those of the inner Galactic sources during their initial ice-forming stages, allowing for the efficient formation of CH$_3$OH.
These points suggest that the Magellanic sources, especially their CH$_3$OH-poor sources, likely experienced environments different from the Galactic ones, thereby affecting the formation efficiency of CH$_3$OH.
\cite{Shimonishi2020} conducted astrochemical simulations to explain the low abundance of CH$_3$OH in LMC hot cores, finding that the CH$_3$OH abundance in a hot core significantly decreases as the visual extinction during the ice-forming stage decreases (see their Figure 9).
This is likely due to the inhibition of hydrogenation reactions of CO under high dust temperatures, and this effect is particularly enhanced in the LMC, where active star formation is ongoing \citep{Shimonishi2020}.
On the other hand, the outer Galactic hot cores likely experienced environments similar to those of the inner Galactic sources during their initial ice-forming stages, allowing for the efficient formation of CH$_3$OH.

\begin{figure*}[tp!]
\begin{center}
\includegraphics[width=18.0cm]{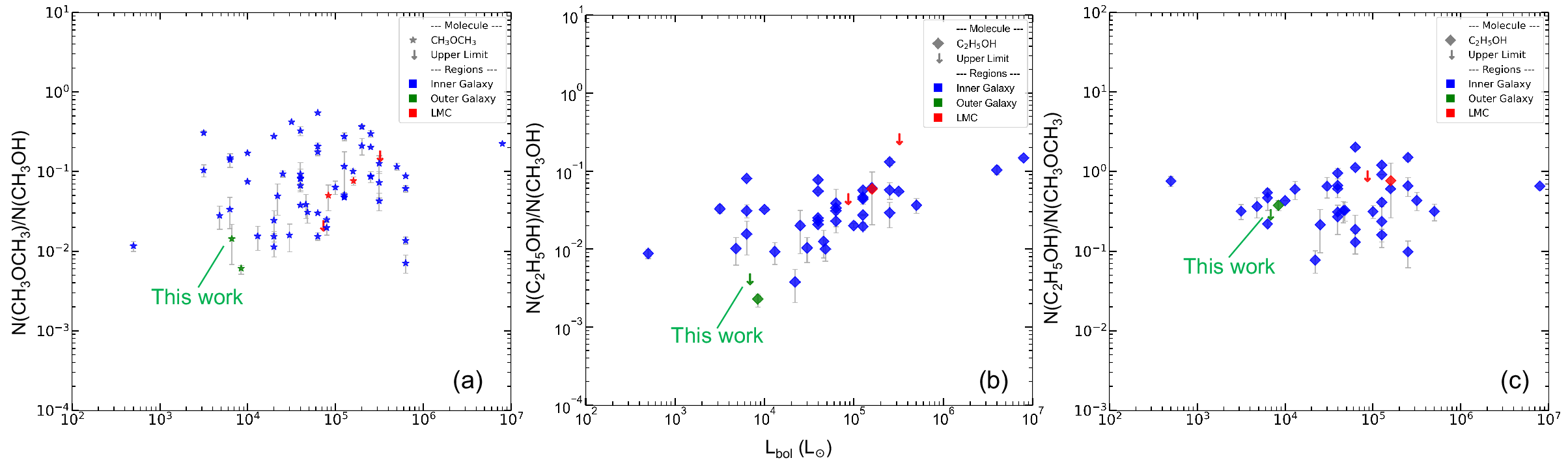}
\caption{
(a) $N$(CH$_3$OCH$_3$)/$N$(CH$_3$OH), (b) $N$(C$_2$H$_5$OH)/$N$(CH$_3$OH), and (c) $N$(C$_2$H$_5$OH)/$N$(CH$_3$OCH$_3$) ratios as a function of luminosity.
The symbols represent inner Galactic hot cores \citep[blue;][]{Fuente2014,Qin2022, chen2023,Li2024, Li2025, Kou2025}, outer Galactic hot cores \citep[green;][]{Shimonishi2021}, and LMC hot cores \citep[red;][]{Shimonishi2016, Shimonishi2020, Broadmeadow2025}.
The subscript arrows in each plot represent the upper limits.
}
\label{im_Abundance_COMs}
\end{center}
\end{figure*}

\begin{figure*}[tp!]
\begin{center}
\includegraphics[width=16.0cm]{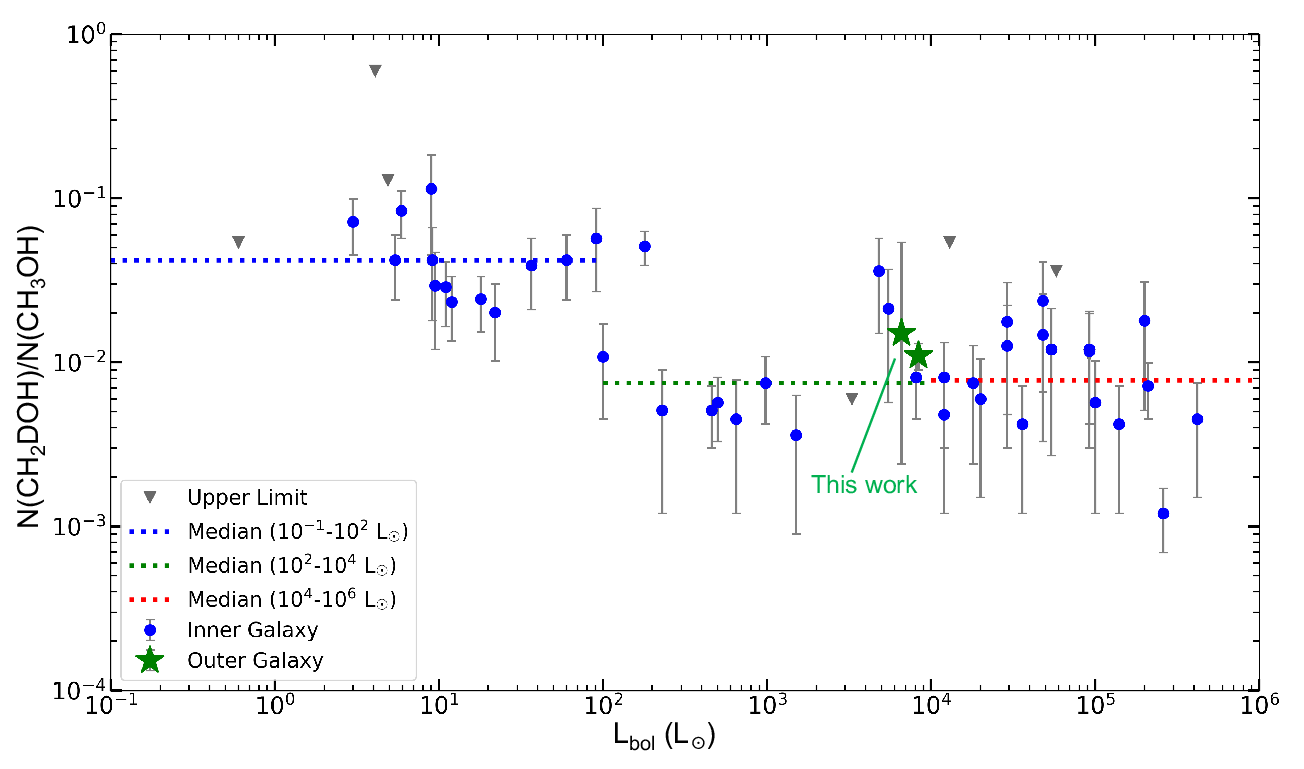}
\caption{The $N$(CH$_2$DOH)/$N$(CH$_3$OH) ratios as a function of luminosities for the inner Galactic sources \citep[blue;][]{Fuente2014,Belloche2016,Muller2016,Jorgensen2018,Taquet2019,Jacobsen2019,Domenech2019,Domenech2021,Manigand2020,Yang2020, van2020,Gelder2022L,Gelder2022,Ligterink2021,Hsu2022, Drozdovskaya2022}, and the outer Galactic sources \citep[green;][]{Shimonishi2021}.
We used the values of the inner Galactic $N$(CH$_2$DOH)/$N$(CH$_3$OH) ratios summarized in \cite{Gelder2022}.
The dotted lines in different color show the median values of $N$(CH$_2$DOH)/$N$(CH$_3$OH) ratio between each luminosity range (blue for 10$^{-1}$ L$_\odot$-10$^{2}$ L$_\odot$, green for 10$^{2}$ L$_\odot$-10$^{4}$ L$_\odot$, and red for 10$^{4}$ L$_\odot$-10$^{6}$ L$_\odot$).
The subscript triangles represent the upper limits.
}
\label{im_Dfraction}
\end{center}
\end{figure*}

Figure \ref{im_Abundance_met_corr}(b) presents a comparison of the $N$(SO$_2$)/$N$(H$_2$) ratios.
This figure indicates that the $N$(SO$_2$)/$N$(H$_2$) ratios of the inner Galactic sources are similar to the metallicity-corrected values of the Magellanic sources.
In contrast, the median value for the outer Galactic sources is significantly lower than that of the inner Galactic sources by approximately a factor of 30.
%Although the number of samples is limited for both the inner and outer Galactic sources, this result would reflect one of the characteristic in the chemical composition of hot cores in the outer Galaxy.
Although SO$_2$ is proposed as a reliable tracer for hot cores in the Magellanic Clouds \citep{Shimonishi2020}, a low SO$_2$ abundance in the outer Galactic hot core has been previously reported \citep{Shimonishi2021}.
%Such a low SO$_2$ abundance is also observed in this work, indicating a characteristic chemical property of hot cores in the outer Galaxy.
Such a low SO$_2$ abundance is also observed in this work, indicating that it might be a characteristic chemical property of hot cores in the outer Galaxy.
SO$_2$ in hot cores is thought to form primarily through solid-phase reactions involving the oxidation of sulfur (S) by radicals (O or OH) or O$_2$ during the early stages of star formation \citep[e.g.,][]{Santos2024}.
Although gas-phase production of SO$_2$ via the successive oxidation reaction of H$_2$S in warm gas also contributes to the observed column densities of SO$_2$ in hot cores \citep[][]{Cha97, Nomura2004}, the SO$_2$ abundance in hot cores would strongly depend on the efficiency of these surface reactions.
It is known that the cosmic ray intensity in the Galactic anticenter region is lower than that in the inner Galaxy \citep{Lipari2025}.
Although the cosmic ray intensity in the outer Galaxy has not been directly constrained (as far as we know), it is expected to be lower than that in the inner Galaxy, considering their results and the lower star formation efficiency in the outer Galaxy.
%\edit1{It is known that cosmic ray intensity is reduced in the outer Galaxy \citep[][]{Bloemen1984}. }
Since cosmic rays play a crucial role in generating radical species on dust grains via cosmic ray-induced UV photodissociation of water ice \citep[e.g., ][and references therein]{Gaches2025}, we speculate that the low SO$_2$ abundance in outer Galactic hot cores results from the insufficient production of oxidizing radicals (O and OH) in its weak cosmic ray environment.

This scenario contrasts with the situation in the Magellanic Clouds.
In these regions, active star formation indicates a high cosmic ray intensity, which ensures an abundant supply of oxidizing radicals via the photodissociation of water ice, thereby enhancing SO$_2$ formation.
Furthermore, unlike the precursor of CH$_3$OH (CO), the precursors of SO$_2$ (S and O) are thought to react within the water-rich ice matrix, which is thermally more stable than the CO-rich ice layer \citep{Santos2024}.
This stability allows these precursors to reside on dust grains and undergo oxidation even under the warmer dust temperatures characteristic of the LMC, where CH$_3$OH formation is inhibited.
Thus, while the CH$_3$OH abundance is critically sensitive to dust temperature during the initial ice-forming stages, the local cosmic ray intensity likely plays a key role in determining the SO$_2$ abundance in hot cores.

\subsection{Molecular abundances: Focusing on COMs} \label{sec_abundance_three} 
Figure \ref{im_Abundance_COMs}(a) and (b) show comparisons of the abundances of COMs (CH$_3$OCH$_3$ and C$_2$H$_5$OH) normalized by CH$_3$OH, and Panel (c) shows a comparison of the $N$(C$_2$H$_5$OH)/$N$(CH$_3$OCH$_3$) ratios among the hot cores in the inner Galaxy (blue), outer Galaxy (green), and LMC (red).
%Blue symbols in each panel represent inner Galactic sources, and the values are extracted from previous observations of inner Galactic hot cores \citep{Qin2022,chen2023,Kou2025,Li2025}.
$D_{\rm{GC}}$ of these inner Galactic sources ranges from $\sim$0.2 to 8 kpc.

Panels (a) and (b) indicate that the $N$(CH$_3$OCH$_3$)/$N$(CH$_3$OH) and $N$(C$_2$H$_5$OH)/$N$(CH$_3$OH) ratios of the inner Galactic and LMC sources are similar over a range of luminosities.
For the outer Galactic sources, the $N$(CH$_3$OCH$_3$)/$N$(CH$_3$OH) ratios are somewhat lower compared to those of the inner Galactic sources ($\sim$1/9 of the median of the inner Galactic ones), lying near the lower limit.
The same applies to the $N$(C$_2$H$_5$OH)/$N$(CH$_3$OH) ratio, as indicated by the other outer Galactic source \citep{Shimonishi2021}.
Note that C$_2$H$_5$OH was not detected in this work, and only an upper limit was derived.

CH$_3$OCH$_3$ and C$_2$H$_5$OH are widely believed to be daughter species of CH$_3$OH, and strong correlations between the column densities of these three species have been confirmed in recent observations of inner Galactic hot cores \citep{Kou2025}.
Based on experimental studies, these molecules are primarily thought to form through the recombination of radicals, which are photodissociation products of CH$_3$OH \citep[e.g., CH$_2$OH, CH$_3$O, and CH$_3$;][]{Oberg2009, Bergantini2018}.
CH$_3$OCH$_3$ is formed by the recombination of CH$_3$ and CH$_3$O radicals \citep{Bergantini2018}:
\begin{equation}
\mathrm{CH}_3 + \mathrm{CH}_3\mathrm{O} \longrightarrow \mathrm{CH}_3\mathrm{OCH}_3.
\end{equation}
C$_2$H$_5$OH is formed by the recombination of CH$_3$ and CH$_2$OH radicals \citep{Bergantini2018}:
\begin{equation}
\mathrm{CH}_3 + \mathrm{CH}_2\mathrm{OH} \longrightarrow \mathrm{C}_2\mathrm{H}_5\mathrm{OH}.
\end{equation}
Considering these formation mechanisms, the lower abundances of these COMs in the outer Galactic sources are likely due to the lower cosmic ray intensity as discussed in Section \ref{sec_abundance_two}, which suppresses the dissociation of CH$_3$OH into their precursor radicals.

On the other hand, the $N$(C$_2$H$_5$OH)/$N$(CH$_3$OCH$_3$) ratios shown in Panel (c) are similar among the sources in the inner Galaxy, outer Galaxy, and LMC.
The values for the outer Galactic and LMC sources (0.38 and 0.77, respectively) are consistent with the median value of the inner Galactic sources (0.43) within a factor of 2.
This remarkable similarity suggests a chemical link between these species, regardless of the differing interstellar environments.
This implies that the formation mechanisms of these COMs in low-metallicity environments are similar to those in the inner Galactic regions, although their formation efficiencies would be somewhat affected by the local environmental conditions (e.g., cosmic ray intensity).
To validate this hypothesis, it is crucial to increase the sample size of COM-rich hot cores in the outer Galaxy and the Magellanic Clouds to enable more robust statistical comparisons.

\subsection{Deuterium fractionation} 
\label{sec_deut}
Deuterium chemistry serves as a crucial tool for deciphering the chemical and physical evolution of interstellar molecules \citep{Cas12, Cec14}.
It is well established that deuterium fractionation occurs efficiently at low temperatures \citep[e.g.,][]{Fur16}.
This efficiency arises from the exothermic nature of the key reaction driving deuterium fractionation \citep[e.g.,][]{Watson1974, Cec14}:
\begin{equation}
\mathrm{H}_{3}^{+} + \mathrm{HD} \longrightarrow \mathrm{H}_2\mathrm{D}^+ + \mathrm{H}_2 + 232\,\mathrm{K}.
\end{equation}
Since the backward reaction is significantly hindered below 20 K, H$_2$D$^+$ is enhanced in cold prestellar cores, leading to an increased atomic D/H ratio in the gas phase.
Moreover, on dust surfaces, deuterium fractionation of CH$_3$OH and related species proceeds effectively under cold conditions via successive H-abstraction and D-substitution reactions \citep[][]{Hidaka2009,Riedel2023}.
These processes result in observed abundances of deuterated molecules (e.g., D$_2$CO and CH$_2$DOH) in star-forming regions that are orders of magnitude higher than the cosmic D/H ratio \citep[(1.52 $\pm$ 0.08) $\times$ 10$^{-5}$;][]{Linsky2003}.
%the formation of deuterated molecules (e.g., D$_2$CO and CH$_2$DOH).

The $N$(CH$_2$DOH)/$N$(CH$_3$OH) ratio of Sh 2-283-1a SMM1 is 1.5$^{+3.9}_{-1.2}$$\%$, which is comparable to that of the other outer Galactic hot core \citep[1.1 $\pm$ 0.2$\%$;][]{Shimonishi2021}.
We also compare the $N$(CH$_2$DOH)/$N$(CH$_3$OH) ratio with those of inner Galactic sources in Figure \ref{im_Dfraction}.
In this figure, we plot the $N$(CH$_2$DOH)/$N$(CH$_3$OH) ratios as a function of luminosity for each source.
Blue symbols represent the values for inner Galactic sources, taken from \cite{Gelder2022}.
The plotted sources are limited to hot cores (or hot corinos) associated with warm gas ($\gtrsim$100 K).

This figure indicates that the $N$(CH$_2$DOH)/$N$(CH$_3$OH) ratios for low-mass sources are higher than those for intermediate- and high-mass sources, while the values for intermediate- and high-mass sources are similar to each other.
The lower $N$(CH$_2$DOH)/$N$(CH$_3$OH) ratios in high-mass sources are generally attributed to their higher temperatures during the prestellar phase and shorter prestellar timescales \citep{Gelder2022, SakaiD2025}.
For Sh 2-283-1a SMM1, the $N$(CH$_2$DOH)/$N$(CH$_3$OH) ratio is comparable to those of intermediate- to high-mass inner Galactic sources.
This result is consistent with its estimated luminosity ($\sim$6.7 $\times$ 10$^{3}$ L$_\odot$).
These results suggest that Sh 2-283-1a SMM1 experienced efficient deuterium fractionation during the formation of CH$_3$OH, which occurred under cold environmental conditions ($\lesssim$ 20 K).

\section{Summary} \label{sec_sum} 
We report the chemical analyses of the outer Galactic hot core, Sh 2-283-1a SMM1 ($D_{\rm{GC}}$ = 15.7 kpc, $Z$ $\sim$0.3 $Z_\odot$), based on ALMA observations.
%Toward this source, a variety of molecular species, including COMs, were detected.
The main conclusions of this study are summarized below.
\begin{enumerate}
\setlength{\itemsep}{1em}
%    \item{A hot core was newly detected in the outer Galaxy with ALMA radio interferometric observations toward the Sh 2-283 region, located at $D_{\rm{GC}}$ = 15.7 kpc (7.9 kpc from the Sun).}

    \item{Toward Sh 2-283-1a SMM1, a variety of molecular species (carbon-, oxygen-, nitrogen-, sulfur-, and silicon-bearing species), including COMs, were detected at the emission peak of 0.87 mm continuum.
    The rotational diagram and continuum analyses revealed the presence of a warm ($T >$100 K), dense ($n \sim$ 4 $\times$ 10$^6$ cm$^{-3}$), and compact ($<$0.03 pc) region associated with the protostar ($\sim$6.7 $\times$ 10$^3$ L$_\odot$).
    }

    \item{A comparison of the molecular abundances relative to CH$_3$OH between the outer Galactic hot cores reveals a similarity, demonstrating that such chemically rich hot cores exist in different regions of the outer Galaxy.
    }

    \item{A comparison of the $N$(CH$_3$OH)/$N$(H$_2$) ratios among the inner Galactic, outer Galactic, and Magellanic hot cores shows that the metallicity-corrected ratios of the outer Galactic hot cores resemble those of the inner Galactic hot cores.
    In contrast, the median value for the Magellanic hot cores is more than one order of magnitude lower than that of the inner Galactic hot cores.
    These results indicate that dust temperature during the initial ice-forming stages significantly affects the CH$_3$OH abundances in hot cores.
    }

    \item{In contrast, the $N$(SO$_2$)/$N$(H$_2$) ratio of the outer Galactic hot cores is significantly lower than that of the inner Galactic hot cores, whereas the values are comparable between the inner Galactic and Magellanic hot cores, indicating that cosmic ray intensity plays a key role in determining the formation efficiency of SO$_2$.
    }

    \item{The $N$(CH$_3$OCH$_3$)/$N$(CH$_3$OH) and \\$N$(C$_2$H$_5$OH)/$N$(CH$_3$OH) ratios of the outer Galactic sources are moderately lower than those of the inner Galactic ones.
    This is likely due to the lower cosmic ray intensity in the outer Galaxy; these COMs are primarily thought to form via radical-radical reactions, and the precursor radicals are produced by the photodissociation of CH$_3$OH induced by cosmic rays.
    On the other hand, the $N$(C$_2$H$_5$OH)/$N$(CH$_3$OCH$_3$) ratios are similar among the inner Galactic, outer Galactic, and LMC sources, indicating the universality of the formation mechanisms of these molecules across the different environments.
    }

    \item{The $N$(CH$_2$DOH)/$N$(CH$_3$OH) ratio of Sh 2-283-1a SMM1 is 1.5$^{+3.9}_{-1.2}$$\%$, which is comparable to that of another outer Galactic hot core and those of inner Galactic intermediate- to high-mass hot cores.
    }
    
\end{enumerate}

\begin{acknowledgments}
This paper makes use of the following ALMA data: ADS/JAO.ALMA$\#$2022.1.01270.S. 
ALMA is a partnership of ESO (representing its member states), NSF (USA) and NINS (Japan), together with NRC (Canada), NSTC and ASIAA (Taiwan), and KASI (Republic of Korea), in cooperation with the Republic of Chile. 
The Joint ALMA Observatory is operated by ESO, AUI/NRAO and NAOJ.
This work has made extensive use of the Cologne Database for Molecular Spectroscopy and the molecular database of the Jet Propulsion Laboratory. 
This work was supported by JSPS KAKENHI grant Nos. JP20H05845, JP21H00037, and JP21H01145. 
This work was supported by the Uchida Energy Science Promotion Foundation. 
%This work was supported by NAOJ ALMA Scientific Research Grant Code 20**-***.
T.I. was supported by the Niigata University Quantum Research Center (NU-Q).
Finally, we would like to thank an anonymous referee for insightful comments, which substantially improved this paper.
\end{acknowledgments}

\appendix
\restartappendixnumbering

\section{The ALMA spectra of the detected molecular emission lines} \label{sec_appen_fit} 
\begin{figure*}[tp!]
\begin{center}
\includegraphics[width=16.0cm]{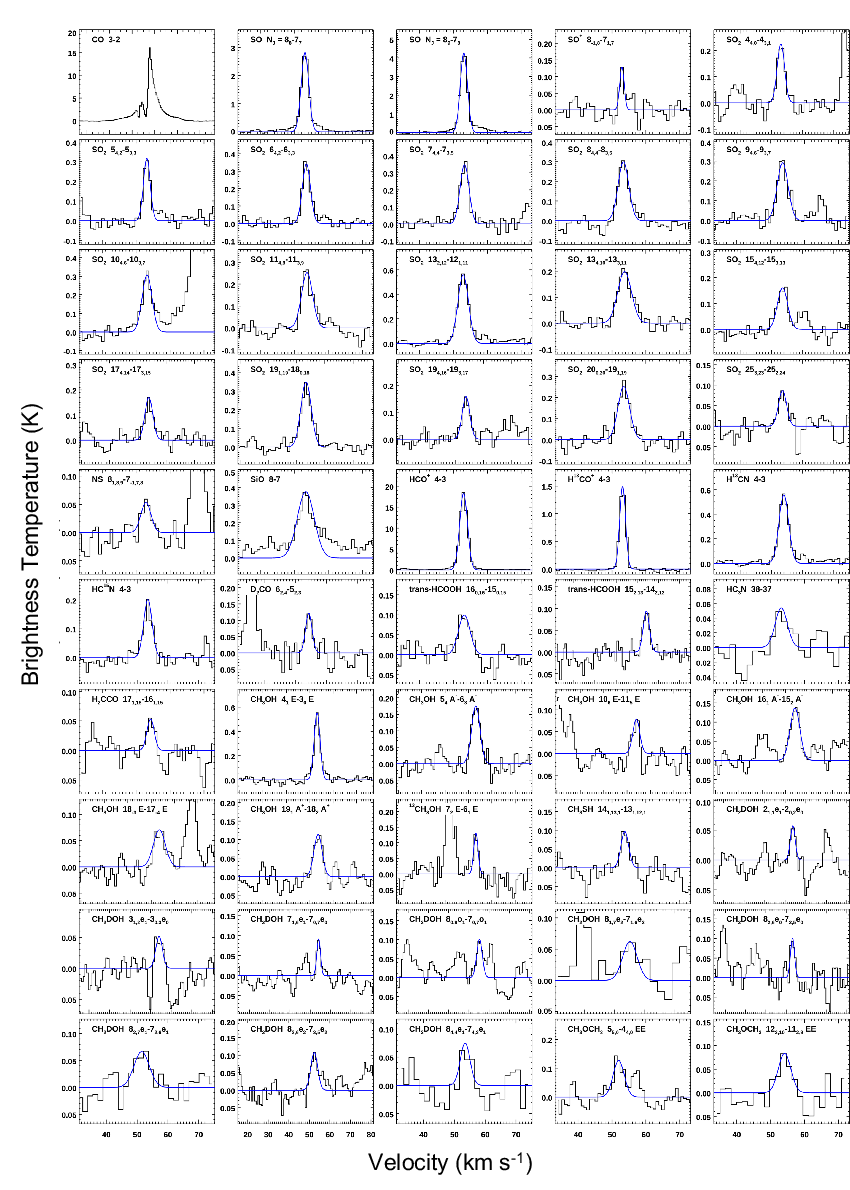}
\caption{ALMA spectra of the detected molecular lines.
The blue lines represent the result of fitting with Gaussian profiles.
For the molecular species which multiple lines are detected, we sort them with the order of the upper state energy.
For the CO($J$ = 3--2) line, the integrated intensity is derived by directly integrating the spectra.
}
\label{im_line_matome}
\end{center}
\end{figure*}

Figure \ref{im_line_matome} shows the ALMA spectra of the detected molecular lines.
The blue solid lines in each spectrum represent the results of the Gaussian fitting.

\section{The molecular abundances of hot cores across the Galaxy} \label{sec_appen_abun} 
\begin{figure*}[tp!]
\begin{center}
\includegraphics[width=18cm]{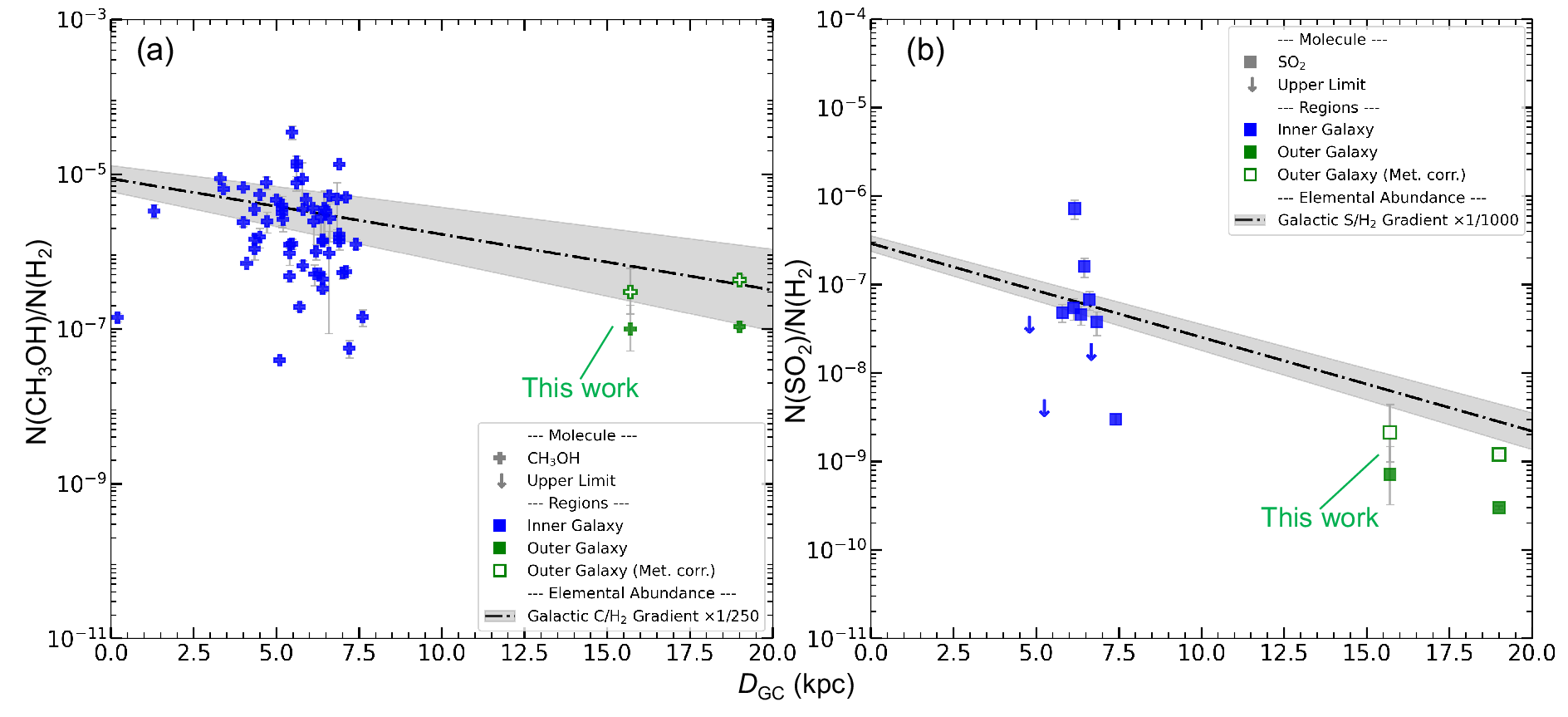}
\caption{The $N$(CH$_3$OH)/$N$(H$_2$) (a) and $N$(SO$_2$)/$N$(H$_2$) (b) ratios as a function of $D_{\rm{GC}}$ for the inner Galaxy sources \citep[blue;][]{Fuente2014,Gelder2022L, Santos2024,Chen2025ATMs,Li2025} and the outer Galactic sources \citep[green;][]{Shimonishi2021}.
The open white marks show the metallicity-corrected values for the low-metallicity sources.
The dotted lines in each panel show the Galactic gradients of the elemental abundances of C \citep{Are20} and S \citep{Fer17} relative to H$_2$.
The intercepts of these gradients are divided by 250 and 1000 for C and S, respectively.
The gray shaded regions in each panel show the error range of the gradients.
}
\label{image_appen_abun}
\end{center}
\end{figure*}

Figure \ref{image_appen_abun} plots the molecular abundances (CH$_3$OH and SO$_2$) as a function of $D_{\rm{GC}}$.
The dotted lines in each panel denote the Galactic gradient of the elemental abundances of C and S measured in H$_\mathrm{II}$ regions \citep{Fer17, Arellano2020}.
The intercepts of these gradients are divided by 250 and 1000 for C and S, respectively.
The plotted Galactic sources are the same as in Figure \ref{im_Abundance_met_corr}.

\end{document}